\documentclass{aa}  
\usepackage[dvipsnames]{xcolor}
\usepackage{hyperref}
\hypersetup{
      colorlinks,
      citecolor=blue,   
      linkcolor=blue,
      urlcolor=blue}
\usepackage{subfigure}
\usepackage{graphicx}
\usepackage{booktabs}
\usepackage{tabularx}
\usepackage{tikz}
\usepackage{orcidlink}
\usepackage{natbib}
\usepackage{textcmds}
\usepackage{mathtools}
\usepackage{sidecap}
\usepackage{multirow}
\usepackage{makecell}

\usepackage{ulem}

\newcommand{\refbold}[1]{#1}
\usepackage[switch]{lineno}

\begin{document}
   \title{Magnetically driven outflows in 3D common-envelope evolution of massive stars}
   \author{
        Marco Vetter\inst{1,2}\orcidlink{0009-0007-2322-6001}
        \and
        Friedrich K. R{\"o}pke\inst{1,2,3}\orcidlink{0000-0002-4460-0097}
          \and 
          Fabian R. N. Schneider\inst{1,2}\orcidlink{0000-0002-5965-1022}
          \and
          R{\"u}diger Pakmor\inst{4}\orcidlink{0000-0003-3308-2420}
          \and
          Sebastian Ohlmann\inst{5}\orcidlink{0000-0002-6999-1725}
          \and
          Javier Mor{\'a}n-Fraile\inst{1,2}\orcidlink{0000-0002-8918-5130}
          \and
          Mike Y. M. Lau\inst{2}\orcidlink{0000-0002-6592-2036}
          \and 
          Giovanni Leidi\inst{2}\orcidlink{0000-0001-7413-7200}
          \and 
          Damien Gagnier\inst{1,2}\orcidlink{0000-0002-1904-2740}
          \and
          Robert Andrassy\inst{1,2}
          }

   \institute{
            Zentrum für Astronomie der Universität Heidelberg, Astronomisches Rechen-Institut, Mönchhofstr. 12-14, D-69120 Heidelberg, Germany\\
            \email{marco.vetter@stud.uni-heidelberg.org}
        \and
            Heidelberger Institut für Theoretische Studien, Schloss-Wolfsbrunnenweg 35, 69118 Heidelberg, Germany\\
        \and
            Zentrum für Astronomie der Universität Heidelberg, Institut für Theoretische Astrophysik, Philosophenweg 12, 69120 Heidelberg, Germany\\
        \and
            Max-Planck-Institut für Astrophysik, Karl-Schwarzschild-Str. 1, D-85748, Garching, Germany\\
        \and
            Max Planck Computing and Data Facility, Gießenbachstraße 2, 85748 Garching, Germany\\
    }
        
   \date{Received YYY; accepted ZZZ}
  \abstract
  {Recent three-dimensional magnetohydrodynamical simulations of the common-envelope interaction revealed the self-consistent formation of bipolar magnetically driven outflows. They are launched from a toroidal structure that has properties of a circumbinary disk.
  So far, the dynamical impact of bipolar outflows on the common-envelope phase remains uncertain and we aim to quantify its importance in this work. Due to the large computational expense of such simulations, we focus on a specific setup to illustrate the impact on common-envelope evolution by comparing two simulations -- one with magnetic fields and one without.
  We use the three-dimensional moving-mesh hydrodynamics code \textsc{arepo} to perform simulations, focusing on the specific case of a $10\, M_\odot$ red supergiant star with a $5\, M_\odot$ black hole companion. 
  We find that by the end of the magnetohydrodynamic simulations (after ${\sim}\, 1220$ orbits of the core binary system), about $6.4\, \mathrm{\%}$ of the envelope mass is ejected through the outflow and contributes to extracting angular momentum from the disk structure and core binary. Given the increased torques induced by the launched material near the core binary, the simulation shows a reduction of the final orbital separation by about $24 \, \mathrm{\%}$ compared to the purely hydrodynamical scenario, while the envelope ejection rate exhibits only temporary differences and is dominated by recombination-driven equatorial winds.
  We further investigate the magnetic-field amplification and the launching mechanism of the bipolar outflows. The results are consistent with previous works: the magnetic fields are primarily amplified by strong shear flows and the magnetically driven outflows are launched by a magneto-centrifugal mechanism. The outflows are additionally supported by local shock heating and strong magnetic gradients and originate from a distance of $1.1$ times the core binary's orbital separation from its center of mass. 
  From this and preceding works, we conclude that the magnetically driven outflows and their role in the common-envelope phase are a universal aspect of such dynamical interactions and we further discuss possible implementations in analytical and non-magnetic numerical model approaches.
  We propose an adaptation of the $\alpha_\mathrm{CE}$-formalism for common-envelope interactions, that accounts for magnetic effects, by modifying the final orbital energy with a factor of $1+ M_\mathrm{out}/\mu$, with $M_\mathrm{out}$ being the mass ejected through the bipolar outflows, and $\mu$ the reduced mass of the core binary.
}

   \keywords{Magnetohydrodynamics --
               Methods:numerical -- 
               Stars:massive -- 
               Stars:supergiants -- 
               binaries:close -- 
               stars:winds,outflows -- 
               stars:magneticfield -- 
               circumstellar material
               }
   \maketitle

\section{Introduction}\label{sec:intro}
The common envelope (CE) phase is one of the most crucial stages in binary stellar evolution. The result of a CE phase has implications for a wide range of astrophysical phenomena, \refbold{including} gravitational wave emitting compact binaries \citep[e.g.,][]{dominik2012a,ivanova2013a,belczynski2016a,tauris2017a,vigna2018a, giacobbo2018a, moreno2022a, roepke2023a}\refbold{, luminous red novae \citep{ivanova2013b, matsumoto2022a,zhuo2024a,hatfull2025a}}, \refbold{and} shaping planetary nebulae \citep[PNe, e.g.,][]{boffin2019a,zou2020a,ondratschek2022a}.
While the fraction of PNe emerging from a preceding CE phase remains unknown, a majority of PNe \citep[$80\, \mathrm{\%} \textit{--} 85 \, \mathrm{\%}$, see][]{sahai1998a,parker2006a} shows morphologies characterized by rapid bipolar outflows and a slowly expanding equatorial waist \citep[bipolar PNe; e.g.,][]{imai2002a, tafoya2020a, guerrero2020a, bollen2020a, bollen2022a}.

As inferred from recent three-dimensional (3D) magnetohydrodynamical (MHD) simulations \citep{ondratschek2022a, vetter2024a}, strongly bipolar structures can self-consistently form in the CE phase. There, the envelope of a giant star is perturbed by a companion plunging deep into the primary star through gravitational drag (e.g., \citealp{paczynski1976a}, \citealp{ivanova2013a}, \citealp{roepke2023a} and \citealp{schneider2024a}). During this in-spiral, the envelope is lifted by the transfer of gravitational energy and angular momentum, leading either to a stable binary formed by the core and the companion (referred to as core binary) or a merging event between the two.
3D simulations of CE evolution have demonstrated that only considering gravitational energy appears to be insufficient to eject the entire envelope \citep[e.g.,][]{ohlmann2016b,chamandy2019a,sand2020a,lau2022a,lau2022b}, and the transferred angular momentum is aligned with the orbital motion of the system. Hence, the formation of a disk-like transient structure around the central core binary after the plunge-in (if the system is not merging) seems inevitable \citep{gagnier2023a} before other energy sources (e.g., release of recombination energy leading to toroidal winds) can further erode the former envelope material \citep{vetter2024a}. Simultaneously, the magnetic fields in the envelope are amplified and drive collimated outflows in the polar directions launched from the disk structure, leaving behind a system with a morphology that is reminiscent of bipolar PNe \citep{ondratschek2022a}. Both disk-like structures and bipolar outflows are formed self-consistently in MHD simulations of different CE models \citep{moreno2022a, ondratschek2022a, vetter2024a}.

Here, we aim to probe this idea through the simulations presented in \citet{vetter2024a}. Furthermore, we quantify the impact of the magnetically driven outflows on the outcome of the CE evolution. While it has been hypothesized to be negligible during the dynamic phase of CE evolution \citep{ondratschek2022a}, we showed in the preceding work \citep{vetter2024a} that the mass transport rates through the bipolar outflows can indeed be as high as $10\, \mathrm{\%}$ of the envelope ejection rate associated with recombination driven winds in the post plunge-in CE system. 

To study these aspects, we proceed in this work as follows: In Sect.~\ref{sec:methods}, we summarize our methods and initial model for this work. We then discuss our results in Sect.~\ref{sec:results}, where we qualitatively compare the MHD simulation to its purely hydrodynamical counterpart (Sect.~\ref{sec:mhd_vs_nomhd}), analyze the magnetic-field amplification (Sect.~\ref{sec:mag_ampl}), investigate the launching mechanisms (Sect.~\ref{sec:mag_launching}) as well as measure the properties of the magnetized outflows with the help of tracer particles (Sect.~\ref{sec:jet_quants}). Furthermore, we compare our results to a CE scenario involving a neutron star companion (NS with $M_2\, {=}\, 1.4\, M_\odot$; Sect.~\ref{sec:ns_comp}). Finally, we continue with a discussion in Sect.~\ref{sec:discussion}, where we set the focus on the implications of magnetic fields in the CE events, and conclude in Sect.~\ref{sec:conclusion}.

\section{Methods}\label{sec:methods}
The models analyzed in this work are the same as those presented by \citet{vetter2024a}, which are based on \citet{moreno2022a}. We briefly summarize key aspects in this work and refer the reader to these publications for further details.

The 3D CE simulations presented in this work are conducted with the moving-mesh MHD code \textsc{arepo} \citep{springel2010a, pakmor2011d, pakmor2013b}. The code employs a second-order finite-volume method for solving the underlying hyperbolic conservation laws. Numerical fluxes are computed at every grid cell boundary with the five-wave Harten--Lax--van Leer--Discontinuities approximate Riemann solver \citep[HLLD,][]{miyoshi2005}. 
The Powell scheme \citep{powell1999a, pakmor2013b} is used to reduce the strength of magnetic monopoles generated during the simulation.
A tree-based algorithm is used to calculate the Newtonian self-gravity \citep{springel2010a}.
Additionally, the new refinement approach presented in \citet{vetter2024a} for the cells within the softened potential around the primary core and companion \citep[see][]{ohlmann2016a} is applied.
To account for recombination energy, we applied the OPAL \citep{iglesias1996a, rogers1996a, rogers2002a} equation of state (EoS) similar to previous CE simulations conducted by, for instance, \citet{sand2020a, kramer2020a} and \citet{moreno2022a}. \refbold{With these choices, we do not account for radiative cooling effects in our models and thus our model is that of an ``adiabatic'' evolution.}

The primary star model for our CE setup is a $10\, M_\odot$ red super-giant (RSG) with a radius of $438\, R_\odot$ as described in \citet{moreno2022a}. The companion is a $5\, M_\odot$ gravity-only point particle chosen to represent a black hole (BH). A dipolar magnetic seed field is applied to the primary star with a surface field strength\footnote{This field strength corresponding to a ratio of gas pressure to magnetic pressure (i.e., plasma-$\beta$) of ${\sim}\, 10^{17}$ at the surface of the red supergiant star. Hence, the magnetic fields are dynamically negligible in the initial setup.} of $1 \, \mathrm{\mu G}$. Additionally, we insert $799,212$ virtual tracer particles (also referred to as \qq{tracers} in the following) as a Lagrangian component recording the physical properties of the flow. The tracer particles are sampled uniformly across the gas cells, such that each represents the same fraction of the total mass of the system. Each tracer particle therefore represents a mass of $m_\mathrm{tr}\, {\approx} \, 8.32 \,{\times}\,10^{-6} \, M_\odot$ \citep[see][]{vetter2024a}. Furthermore, we conducted a simulation that has the same setup but without magnetic fields, which we refer to as the \qq{purely hydrodynamical} simulation (also \qq{Hydro} case).

The mass of unbound material is quantified by summing over all cells with positive specific energy, where we distinguish between
\begin{align}
    e_\mathrm{kin} + e_\mathrm{pot} > 0 , \;&\textrm{``kinetic-energy criterion''}, \label{eq:e_kin_crit}\\
    e_\mathrm{kin} + e_\mathrm{therm} + e_\mathrm{pot} > 0 , \; &\textrm{``thermal-energy criterion'',} \label{eq:e_therm_crit}\\
    e_\mathrm{kin}+ e_\mathrm{int} + e_\mathrm{pot} > 0 , \; &\textrm{``internal-energy criterion''},\label{eq:e_int_crit}
\end{align}
with $e_\mathrm{kin}$, $e_\mathrm{therm}$, $e_\mathrm{int}$ and $e_\mathrm{pot}$ being the specific kinetic, thermal, internal (i.e., the sum of the thermal and the ionization energies) and potential energy\footnote{ 
Including magnetic energy yields the same result because all other energy density contributions dominate locally.}.

\begin{figure}
    \includegraphics[width=\linewidth]{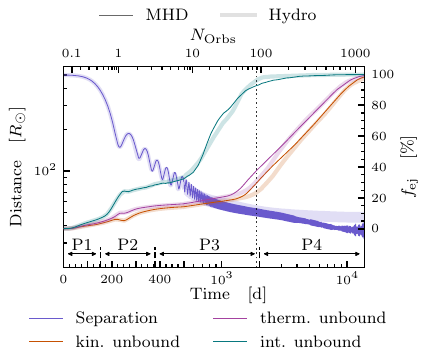}
    \caption{Evolution of the orbital distance and envelope ejection fraction $f_\mathrm{ej}$ for the MHD (thin lines) and Hydro case (thick lines). Shown are the orbital separation (blue) and the fraction of unbound material according to the kinetic, thermal and internal energy criteria in orange, purple and green, respectively. The black dotted vertical line indicates the starting time of the magnetically driven outflows and the labels \qq{P1} to \qq{P4} represent the different magnetic-field amplification phases identified in Sect.~\ref{sec:mag_ampl}.}
    \label{fig:MHD_vs_noMHD_oe_fej}
\end{figure}

\section{Results}\label{sec:results}
\subsection{MHD versus Hydro}\label{sec:mhd_vs_nomhd}

\begin{figure}
    \includegraphics[width=\linewidth]{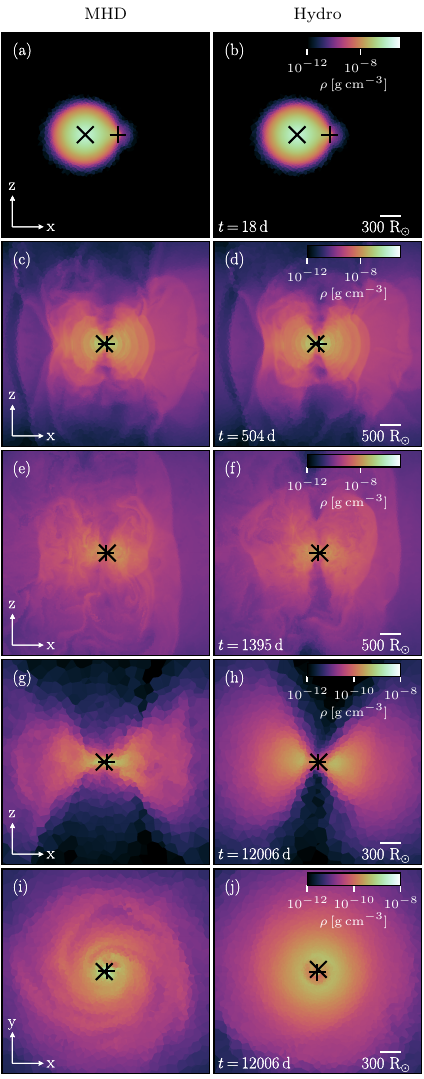}
    \caption{Temporal evolution of the density in the MHD simulation (left) and Hydro simulation (right). Each row shares the same time, orientation, color scale and box size. Panel (a) to (h) are edge-on ($x$-$z$) views, while (i) and (j) show the system face-on ($x$-$y$). The primary core and companion are marked with a black \qq{x} and \qq{+}, respectively.}
    \label{fig:MHD_vs_noMHD}
\end{figure}

\begin{figure*}
  \begin{minipage}[c]{0.7\textwidth}
        \includegraphics[width=\textwidth]{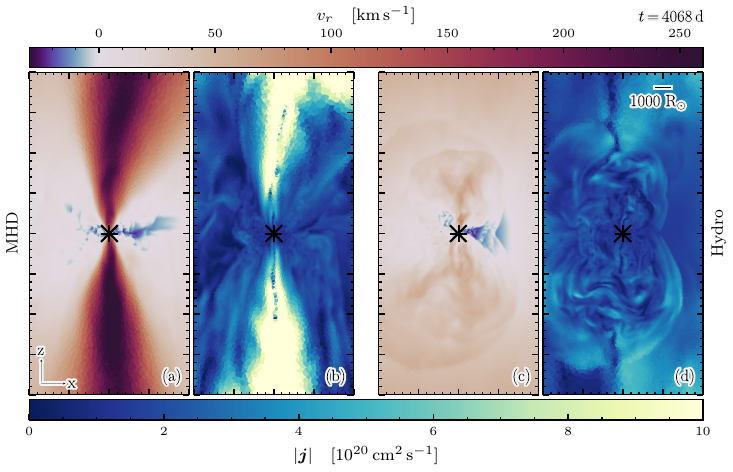}
  \end{minipage}\hfill
  \begin{minipage}[c]{0.27\textwidth}
        \caption{Comparison between the MHD (two plots on the left) and Hydro simulation (two plots on the right) in edge-on slices of radial velocity $v_r$ ([a] and [c]) and absolute specific angular momentum $\vert \boldsymbol{j} \vert$ ([b] and [d]) with respect to the center of mass of the core binary at $t \, {=}\, 4068\, \mathrm{d}$. The MHD simulation shows faster bipolar outflows that carry away a higher specific angular momentum compared to the Hydro case.} 
        \label{fig:vr_j}
  \end{minipage}
\end{figure*}

For the MHD and the Hydro model alike, as soon as the companion is placed at the surface of the primary star, it starts plunging deeply into the envelope within the first few orbits (Fig.~\ref{fig:MHD_vs_noMHD_oe_fej} for $t \, {\lesssim}\, 500 \, \mathrm{d}$). During this inspiralling motion, the density perturbations in the envelope are nearly indistinguishable in the two simulations (cf., Figs.~\ref{fig:MHD_vs_noMHD}a and~\ref{fig:MHD_vs_noMHD}b, and Figs.~\ref{fig:MHD_vs_noMHD}c and~\ref{fig:MHD_vs_noMHD}d) and, consequently, we find a good agreement in the orbital evolution and envelope ejection (Fig.~\ref{fig:MHD_vs_noMHD_oe_fej}). 
As soon as the systems evolve further, beyond the rapid spiral-in (Fig.~\ref{fig:MHD_vs_noMHD}a, for $t \, {\gtrsim}\, 500 \, \mathrm{d}$), we start to observe deviations not only in the orbital separation but also in the unbound mass fractions.

Although the deviations in orbital separation are small after the in-spiral, they become increasingly larger and the orbital separation averaged over the last 20 orbits of the core binary increased from $37$ to $46 \, R_\odot$ (by ${\approx} \, 24 \, \mathrm{\%}$) in the Hydro case compared to the MHD simulation.
Since the core binary can only interact gravitationally with the ambient medium in the simulation, this finding can only be explained by increased gravitational torques in the MHD simulation acting on the binary.
Comparing the post spiral-in density structures of the simulations (Fig.~\ref{fig:MHD_vs_noMHD} and also Movie M1 in Table~\ref{tab:movietable}), it becomes evident that the Hydro case appears less turbulent and more symmetric. Even a clear cavity and evacuated bipolar funnels can be observed in the proximity of the core binary (Fig.~\ref{fig:MHD_vs_noMHD}).
The turbulent structures may be traced back to increased stresses by magnetic fields \citep{gagnier2024a} and in fact, from $200 \, \mathrm{d}$ on, the magnetic fields become strong enough to extract kinetic energy from the gas (see Appendix~\ref{sec:appendixA}, Fig.~\ref{fig:ekin_evolution}). Regardless, a more detailed analysis is required to shed light on the complex mechanism behind the angular momentum transport.

We now study the differences in the evolution of the amount of unbound mass. In both runs, we find recombination-driven winds that dominate the overall envelope ejection process \citep[see also][]{vetter2024a}; they can be most easily identified by the rapid increase in ejected material according to Eq.~(\ref{eq:e_kin_crit}) at $t\, {\approx}\, 1400\,\mathrm{d}$ in Fig.~\ref{fig:MHD_vs_noMHD_oe_fej}. In the MHD case, these winds appear about $500\, \mathrm{d}$ earlier compared to the Hydro case (see the orange lines in Fig.~\ref{fig:MHD_vs_noMHD_oe_fej}).
One might first think of the amplified magnetic fields as a potential explanation for this finding (see Sect.~\ref{sec:mag_ampl}). But in fact, the total magnetic energy of the entire envelope at about $1400 \, \mathrm{d}$ is approximately two orders of magnitude below the kinetic energy and furthermore the amplification of the magnetic fields acts as an energy drain, which is the reverse trend of what is needed to explain the earlier onset of the recombination-driven winds. 
However, there is one energy source that is increasingly contributing to the kinetic energy budget of the system in the MHD case: the larger released orbital energy. In fact, the difference in (averaged) orbital energy between the Hydro and the MHD case at this point is about $2\times10^{46}\, \mathrm{erg}$ and thus approximately comparable to the difference in total kinetic energy between the two simulations ($E_\mathrm{kin, \, MHD} - E_\mathrm{kin, \, Hydro} \, {\approx} \, 10^{45}\, \mathrm{erg}$ at $1400\, \mathrm{d}$ and up to ${\approx} \, 10^{46}\, \mathrm{erg}$ at $2000\, \mathrm{d}$). We thus conclude that the earlier onset of recombination in the MHD case can be traced back to the larger release of orbital energy by the core binary. 
As the evolution continues, these differences become negligible as the simulations evolve towards full envelope ejection, and the unbound mass fraction follows a similar evolution in both cases for $t\,{\gtrsim} \, 4000 \, \mathrm{d}$. However, the most striking difference between the runs is that the MHD model develops high-velocity, collimated, bipolar outflows (see Movie M2 in Table~\ref{tab:movietable}) which carry away angular momentum (Fig.~\ref{fig:vr_j}), as we will show in Sect.~\ref{sec:jet_quants} \citep[see][]{vetter2024a}.
\subsection{Magnetic-field amplification}\label{sec:mag_ampl}

\begin{figure*}[h!]
    \includegraphics[width=\textwidth]{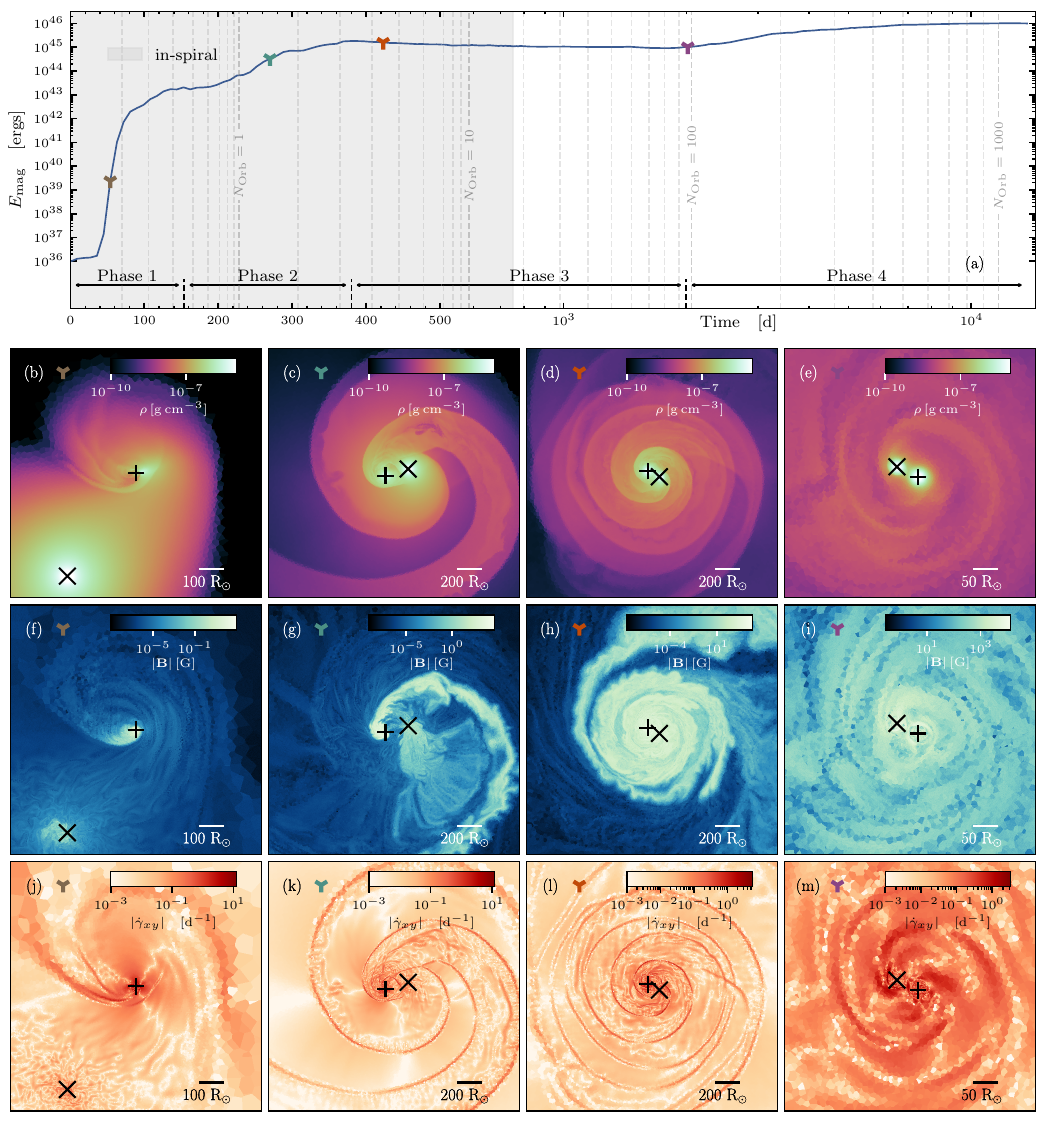}
    \caption{Temporal evolution of the total magnetic energy in panel (a). The labels \qq{Phase 1} to \qq{Phase 4} are for the different magnetic-field amplification phases observed in our simulation (see Sect.~\ref{sec:mag_ampl}) and the grey-shaded region highlights the in-spiral phase defined as the time for which $\vert \dot{a}/a\vert  P > 0.05$, with $a$ being the orbital separation and $P$ being the orbital period. In panels (b)-(e), (f)-(i) and (j)-(m), we show the density $\rho$, the absolute value of the magnetic field $\vert \boldsymbol{B} \vert$ and the absolute value of the local shear rate in the orbital plane $\vert \dot{\gamma}_{xy} \vert = \vert \partial_y v_x + \partial_x v_y \vert$, respectively. 
    The colored markers in panels (b)-(m) are the same as those in panel (a) and indicate the time at which the quantities are shown. The primary core and companion are again marked by a \qq{x} and \qq{+} symbol, respectively.}
    \label{fig:mag_ampli}
\end{figure*}

As already indicated in \citet[][]{vetter2024a}, we observe strong amplification of the seed magnetic fields in our CE simulation, in agreement with \citet{ohlmann2016b} and \citet{ondratschek2022a}. We find an exponential increase in the total magnetic energy by roughly 10 orders of magnitude during multiple distinct amplification phases (Fig.~\ref{fig:mag_ampli}a).

The first phase takes place during the first orbit up to ${\sim} \, 154 $ days (see Fig.~\ref{fig:mag_ampli}a, the time span labeled as \qq{Phase 1}). It is characterized by the supersonic motion of the companion in the initially only weakly magnetized envelope material within the first orbit \citep[corresponding to the ``fast amplification'' phase in][]{ohlmann2016b}. 
During this phase, the initial perturbation of the envelope creates an accretion stream towards the companion. The spun-up gas in the vicinity of the companion gives rise to locally enhanced shear with respect to the still unperturbed envelope that leads to an amplification of the seed magnetic field (Fig.~\ref{fig:mag_ampli}b, \ref{fig:mag_ampli}f and \ref{fig:mag_ampli}j).
The expected growth time of the magneto-rotational instability \citep[MRI,][]{balbus1991a} in the fast amplification phase is in our case as short as $2\, \textrm{--}\, 4\,\mathrm{d}$ and approximately matches the e-folding growth time in the magnetic energy of roughly $4\, \mathrm{d}$. In our calculation of the MRI growth rate $\gamma_\mathrm{MRI}$ and growth time $\tau_\mathrm{MRI} \, {=}\, \gamma_\mathrm{MRI}^{-1}$,  we follow \citet{rembiasz2016a}, who give $\gamma_\mathrm{MRI} \, {=} \,  q\Omega/2$, where $q$ is the local rotational shear and $\Omega$ is the angular velocity. The first phase is terminated as soon as the companion completed half an orbit and plunged deeply into the envelope of the primary star. At this point in time, the orbital contraction starts to slow (see Fig.~\ref{fig:MHD_vs_noMHD_oe_fej}).

In the second phase (\qq{Phase 2} in Fig.~\ref{fig:mag_ampli}a), the spiral-shock arm created within the first orbit (Fig.~\ref{fig:mag_ampli}c) expands, becomes unstable, and triggers amplification (Fig.~\ref{fig:mag_ampli}g). Along the entire structure, we observe increased shear rates (Fig.~\ref{fig:mag_ampli}k), which most likely give rise to Kelvin--Helmholtz instabilities.
These phenomena act until ${\approx}\, 380\, \mathrm{d}$ and dominate the second amplification phase \citep[the slow amplification phase in][]{ohlmann2016a}. The expected growth time of the MRI in this phase reaches $90\, \textrm{--}\, 121\,\mathrm{d}$, which is comparable to the e-folding growth time in Fig.~\ref{fig:mag_ampli}a of ${\sim}\, 97\, \mathrm{d}$ and thus supports the idea of the MRI being triggered once more as a secular instability superimposed on the shear instability.

Similar to \citet{ondratschek2022a}, we observe one additional weak dynamo at play, which is present in the first amplification phase and in the beginning of the second. There, the fluid motion near the primary core leads to an amplified magnetic-field strength (Fig.~\ref{fig:mag_ampli}f near the core of the primary star).

Although the companion and primary core are still perturbing the remaining envelope and imprinting a spiral pattern in the density profile (e.g., Figs.~\ref{fig:mag_ampli}d and \ref{fig:MHD_vs_noMHD}c), the second amplification phase is terminated after approximately three orbits ($\approx 380 \mathrm{d}$) and the magnetic energy reaches saturation. In fact, the magnetic energy in the bound material reaches ${\approx} \, 1\,\textit{--}\,2\, \%$ of the total kinetic energy, and we reach equipartition values between the magnetic and turbulent kinetic energy\footnote{For the range in values, we perform volume integrals of the energy densities over a sphere with radius $r\,{\in}\, [100,\, 200,\, 300,\, 400,\, 500,\, 1000] \,R_\odot$ centered on the center of mass of the core binary. The equipartition values are computed under the assumption that the turbulent velocity of the gas can be approximated as $\boldsymbol{v}_\mathrm{turb} \, {=} \, (v_r, v_z)^T$, where $v_r$, $v_z$ are the radial and vertical velocity components in cylindrical coordinates.} of about $ 30\,\textit{--}\,40\, \%$. In this phase, the magnetic energy starts to decrease on timescales on the order of ${\sim}\, 10^{4}\, \mathrm{d}$ (marked as \qq{Phase 3} in Fig.~\ref{fig:mag_ampli}a).

Phase~3 of decreasing magnetic-field strength is followed by the third and final amplification phase at around $t\, {\sim}\, 2000 \, \mathrm{d}$ (labelled \qq{Phase 4} in Fig.~\ref{fig:mag_ampli}a).
While the requirements are met for the MRI to operate in the bound material ($\mathrm{d} \Omega^2 /\mathrm{d} r \, {<}\, 0 $, see \citealp{vetter2024a}), the magnetic-energy growth reaches e-folding times of ${\approx}\, 2500\, \mathrm{d}$, which exceed the expected growth time of the linear MRI by an order of magnitude. We also expect linear MRI to be suppressed because the kinetic and magnetic energy densities are comparable at the onset of Phase~4 in the bound material.

Nonetheless, we observe increased shear rates near the core binary (Fig.~\ref{fig:mag_ampli}m) which induce further growth in the magnetic energy. The magnetic energy reaches values of about $30 \textit{--}90\, \%$ of the volume-integrated kinetic energy and even super-equipartion between $1\, \textit{--}\, 5$ times the turbulent kinetic energy. Hence, the magnetic energy is no longer sourced by turbulent kinetic energy but also by the kinetic energy of the differentially rotating fluid around the core binary. \citet{ondratschek2022a} proposed that this phase may be originating from shear flows induced by mass transfer in the core binary, similar to the case of a main-sequence or WD-NS merger event \citet{schneider2019a,moran2024a}.
Further analysis is required to better understand the last magnetic-field amplification phase.

To deepen our understanding of the temporal evolution of the magnetic-field amplification, we turn our attention to the evolution equation of the magnetic energy density in ideal MHD, 
\begin{align}
    \partial_t e_\mathrm{mag} + \boldsymbol{\nabla} \cdot (e_\mathrm{mag} \boldsymbol{v})= \boldsymbol{B} \cdot (\boldsymbol{B}\cdot \nabla)\boldsymbol{v} - e_\mathrm{mag}\boldsymbol{\nabla} \cdot \boldsymbol{v},\label{eq:emag_eq}
\end{align}
where $\boldsymbol{B}$ is the magnetic field, $e_\mathrm{mag} = \boldsymbol{\vert B\vert^2}/2$ is the magnetic energy density and $\boldsymbol{v}$ is the velocity.
Here, the two terms on the left-hand side of the equation correspond to the temporal change and the transport of magnetic energy in conservation form, while the terms on the right-hand side are associated with stretching and (de-)compression of field lines. 

\begin{figure}
    \includegraphics[width=\linewidth]{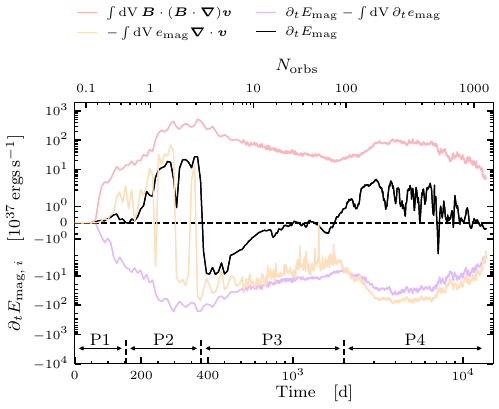}
    \caption{Time evolution of the different terms in the magnetic energy density evolution equation (see Sect.~\ref{sec:mag_ampl}, Eq.~\ref{eq:emag_eq}) integrated over the entire domain. Shown are the stretching term $\int \mathrm {d}V\, \boldsymbol{B} \cdot (\boldsymbol{B} \cdot \boldsymbol{\nabla} )\boldsymbol{v}$, compression term $\int \mathrm{d}V\,  e_\mathrm{mag} \boldsymbol{\nabla \cdot v}$, the time derivative of the total magnetic energy $\partial_t\,  E_\mathrm{mag}$ and the numerical dissipation term ($\partial_t \, E_\mathrm{mag} - \int \mathrm{dV} \, \partial_t \, e_\mathrm{mag}$) in red, orange, black and violet, respectively. Given the applied periodic boundary condition in our model, the contribution of the advection term vanishes, and for the analysis here we focus on the compression and stretching terms. The time derivative of the magnetic energy $\partial_t E_\mathrm{mag}$ is obtained from the data in Fig.~\ref{fig:mag_ampli}a.}
    \label{fig:emag_evolution}
\end{figure}

Integrating the individual terms over the entire domain yields the global evolution of the magnetic energy, as shown in Fig.~\ref{fig:emag_evolution}. Again, we can identify the four phases indicated previously in Fig.~\ref{fig:mag_ampli}.
We find that the stretching term dominates the field amplification throughout the entire simulation \citep[similar to the findings in][]{gagnier2024a}. The highest local shear rates (and since $\boldsymbol{B} \cdot (\boldsymbol{B}\cdot \nabla)\boldsymbol{v} \, {=}\, 0.5 B_iB_j\dot{\gamma}_{ij}$, the fastest amplification; compare with Fig.~\ref{fig:mag_ampli}) are found within the spiral arms and near the companion and primary core.
Given the strong shock initiated in the perturbed envelope by the plunge-in and the emerging spiral structure afterward, we observe the compression term contributing to the first and partially to the second amplification phase via shock compression of the material. However, during the second half of the second phase (around ${\approx}\, 300 \, \mathrm{d}$), the compression term becomes negative due to the expanding envelope material, and the amplification is terminated. Similar to the first two phases, the stretching term increases again in the fourth phase, most likely caused by strong differential rotation near the core binary and close to the spiral pattern as argued before.

\subsection{Magnetic outflow launching}\label{sec:mag_launching}
\begin{figure*}
  \begin{minipage}[c]{0.7\textwidth}
        \includegraphics[width=\textwidth]{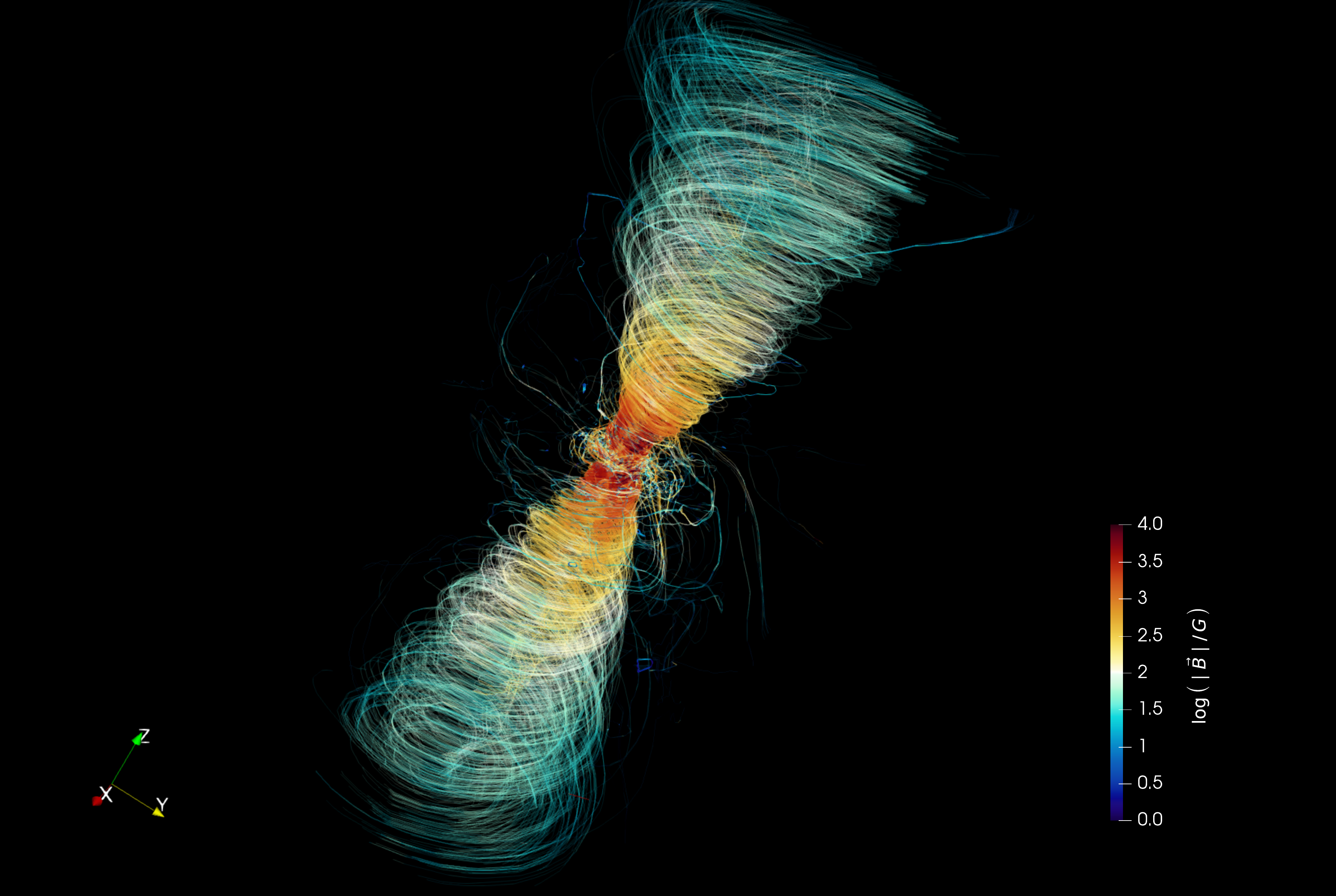}
  \end{minipage}\hfill
  \begin{minipage}[c]{0.27\textwidth}
    \caption{3D rendering of the magnetic-field structure at $t\, {=}\, 4950\, \mathrm{d}$. The color represents the absolute magnetic-field strength. The core binary is embedded between the two conical structures, and the orbital plane is the $x$--$y$ plane.}
    \label{fig:overview_bfield}
  \end{minipage}
\end{figure*}
\begin{figure}      
    \includegraphics[width=1.0\linewidth]{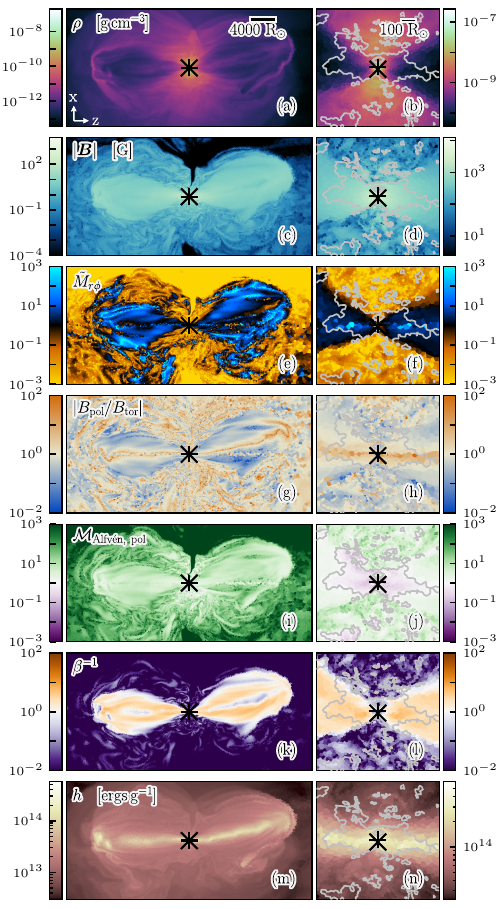}
    \caption{Properties of the bipolar outflows at $t\, {=} \, 3510 \,\mathrm{d}$. Shown are the density $\rho$ in (a) and (b), the absolute value of the magnetic field $\vert \boldsymbol{B} \vert$ in (c) and (d), the $r\phi$-component of the Maxwell stress tensor normalized by the centrifugal force $\tilde{M}_{r\phi}\, {=}\, B_rB_\phi/4\pi\rho v_\phi^2$ in (e) and (f), the absolute ratio between the poloidal and toroidal magnetic-field components $\vert B_\mathrm{pol}/B_\mathrm{tor}\vert$ in (g) and (h), the poloidal Alfvén Mach number $\mathcal{M}_\mathrm{Alfven, pol}=v_\mathrm{pol}\sqrt{4\pi \rho}/B_\mathrm{pol}$ in (i) and (j), the inverse plasma beta $\beta^{-1} = |\boldsymbol{B}|^2/8\pi P_\mathrm{gas}$ in (k) and (l), as well as the specific enthalpy $h$ in (m) and (n). The gray contours in the zoomed-in plots in the left column correspond to the transition from sub- to super Alfvénic flows (i.e., the Alfvén surface).}
    \label{fig:magn_launching}
\end{figure}

Once the central binary enters the third amplification phase (Phase 4 in Fig.~\ref{fig:mag_ampli}a, around $2000\, \mathrm{d}$ or $100$ orbits; cf., Sect.~\ref{sec:mag_ampl}), a magnetically driven outflow perpendicular to the orbital plane becomes visible (e.g., Fig.~\ref{fig:vr_j}). As already stated in Sect.~\ref{sec:mhd_vs_nomhd}, we find along the chimney, that is the low density funnel along the $z$-direction, outflows with enhanced radial velocities carrying away angular momentum and mass (see Fig.~\ref{fig:magn_launching}a and \ref{fig:magn_launching}b). 
While in the Hydro case, the outflows exhibit lower velocities and are less persistent over time (even transitioning between unipolar and bipolar; see Movie M2 in Table~\ref{tab:movietable}), the amplified fields in the MHD run boost the acceleration, lead to collimation and create persistent polar outflows.
In the Hydro case, the outflows may be understood as local shock heating of material near the core binary pushing gas out of the gravitational potential along the way of least resistance, that is, in the polar direction (\citealp{narayan1995a}, similar to the findings in \citealp{gagnier2024b} and potentially in \citealp{lau2022b}). A thermally driven acceleration like that leads to increased entropy and specific enthalpy in the outflows, which indeed is observed in both our simulations \citep[Fig.~\ref{fig:magn_launching}m and \ref{fig:magn_launching}n for the enthalpy and for the entropy we refer to][]{vetter2024a}. However, given the temporal robustness and strength of the bipolar outflow in the MHD case, the acceleration by local shock heating described in \citet{gagnier2024b} can be a contribution, but not the main driving mechanism of the acceleration in our case.

Within the Alfvén surface (the transition between sub- to super-Alfvénic poloidal speeds, marked as a gray contour in the zoomed-in plots on the right-hand side of Fig.~\ref{fig:magn_launching}), the gas is accelerated in the polar direction and the low Alfvén Mach numbers $\mathcal{M}_\mathrm{A} \, {=}\, v_\mathrm{pol}\sqrt{4\pi \rho}/B_\mathrm{pol}$ are accompanied by magnetically dominated flows ($\beta^{-1} \,{=}\, \boldsymbol{B}^2/(8\pi P_\mathrm{gas})\, {>}\, 1$, see Fig.~\ref{fig:magn_launching}k and \ref{fig:magn_launching}l). There, we additionally observe the Maxwell stresses dominating over the centrifugal force $B_rB_\phi/(4\pi \rho v^2_\phi) \, {>}1$ (Fig.~\ref{fig:magn_launching}e and f) indicating that the magnetic fields govern both the gas’s rotation and the redistribution of angular momentum in the bipolar outflow \citep{garcia2021a}. Consequently, the magnetic-field lines are wound up in the outflows, and we observe the helical structure (Fig.~\ref{fig:overview_bfield}) as expected in a magneto-centrifugally driven ejection, such as in the Blandford--Payne mechanism \citep{blandford1982a}.
Furthermore, and in agreement with \citet{ondratschek2022a}, we also observe a magnetic field dominated by its poloidal component (Fig.~\ref{fig:magn_launching}g and \ref{fig:magn_launching}h) near the central binary and low poloidal Alfvén Mach numbers $\mathcal{M}_\mathrm{A}\, \, {<} \, 1$ in the regions above and below the orbital plane of the central binary (Fig.~\ref{fig:magn_launching}j). Both are again characteristics of the Blandford--Payne mechanism for driving magnetic jets, and this process contributes to the outflow of material in our setup. We further identify regions dominated by the toroidal component of the magnetic field, primarily in the circumbinary material (Fig.~\ref{fig:magn_launching}h) resulting from differential rotation. This effect, even though not as pronounced as in \citet[][]{ondratschek2022a}, also contributes to the acceleration, similar to magnetic tower jets \citep[e.g.,][]{uchida1985a, lynden1994}. 

The cumulative effect of all mechanisms is to accelerate the gas in the outflowing regions to a maximum radial velocity of $v_r \, {=}\, 290\,\mathrm{km\,s^{-1}}$, which is of the same order as the orbital velocity of the central binary ($v_\mathrm{orb}\, {=}\, \sqrt{GM_b/a}\, {\approx}\, 175\, \textit{--}\, 205 \,\mathrm{km\,s^{-1}}$ for $t\,{=}\,2000\,\textit{--}\, 13,708\,\mathrm{d}$, with $M_b$ and $a$ being the core binary mass and its orbital separation, respectively). The half-opening angle of the funnel varies within ${\approx}\,  8\, {\textit{--}}\, 12^\circ$, where we determined the opening angle according to the gas cells with velocities exceeding $v_r\, {=} \, 110\, \mathrm{km\, s^{-1}}$. Above the Alfvén surface, we observe supersonic ($\mathrm{\mathcal{M}}\, {>}\, 1$) and super magneto-sonic speeds ($\mathcal{M}_\mathrm{ms} \, {=} \, \vert \boldsymbol{v} \vert / \sqrt{v_\mathrm{A}^2 + c_s^2} \, {>}\, 1$, with $v_\mathrm{A}\, {=}\, (\vert\boldsymbol{B}\vert^2/\sqrt{4\pi \rho})^{1/2}$ being  the Alfvén velocity), progressively increasing with outflowing velocity, that is, the highest radial velocities are observed around the symmetry axis of the outflow, and they decrease with distance to the axis. 

\subsection{Drifting with the flow: Properties of the bipolar outflow}\label{sec:jet_quants}

In the following, we use the implemented tracer particles (see Sect.~\ref{sec:methods}) to analyze the magnetically driven outflow. A tracer particle is considered to be launched by the magnetically driven outflow if, at some point during the simulation, its radial velocity exceeds a critical threshold\footnote{With this approach, the post-processing of the tracer particles is sensitive to the actual choice of the critical parameter and provides only an order-of-magnitude evaluation of the quantities carried away by the gas, rather than a precise measurement.}, $v_{r,\, \mathrm{crit}}$. 
We show results of our analysis assuming different thresholds of $v_{r,\,\mathrm{crit}} \, {=} \, 110,\, 160,\, 210\,\mathrm{km\,s^{-1}}$. These choices include the broader outer region of the outflow, the intermediate accelerated areas and the very inner region of the outflowing material, respectively, but exclude material accelerated in the bow shocks, which typically has velocities below $110\, \mathrm{km\, s^{-1}}$. We note that the smallest considered $v_{r,\, \mathrm{crit}}$ already exceeds the escape velocity from the inner binary at a distance of ${\approx}\,  250 \, R_\odot$ (coinciding with the typical distances spanned by the Alfvén-pocket in Fig.~\ref{fig:magn_launching}, gray iso-surfaces) and the typical velocities in the recombination-driven winds of about ${\sim} \, 50\, \mathrm{km\, s^{-1}}$.
For the three criteria, we find a total of $51,077$, $30,811$ and $12,649$ tracer particles to be ejected in the polar directions throughout our simulation.

\subsubsection{Origin and trajectory}\label{sec:tr_origin}

Given our selected tracer particles in the magnetically driven outflows, we can reconstruct their origin in the stellar structure and trajectory. To obtain the broadest possible overview, we take all tracer particles into account which are found to be launched according to the least constraining condition for the radial velocity (i.e., all tracer particles which exceed $v_r\, {=}\, 110\, \mathrm{km\,s^{-1}}$, see Sect.~\ref{sec:jet_quants}).

\begin{figure}
    \centering
    \includegraphics[width=\linewidth]{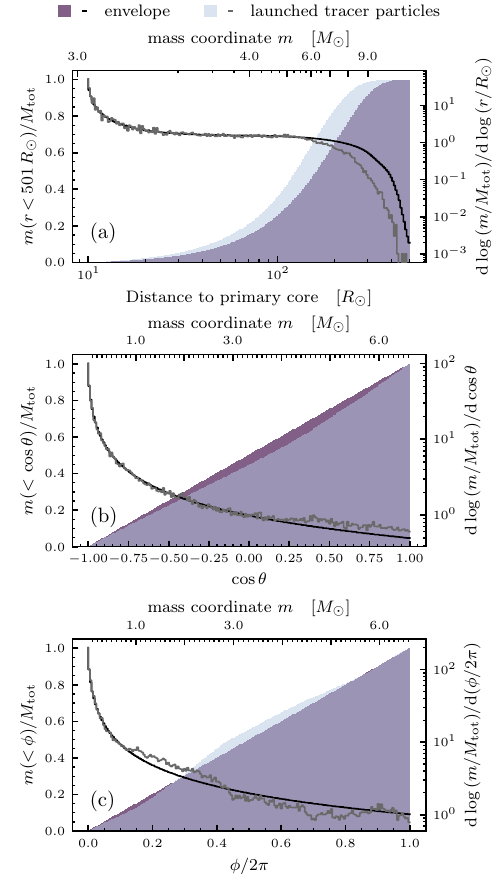}
    \caption
	{Initial distribution of the launched tracer particles within the envelope. Shown are the normalized cumulative mass distributions $m({<}\, x)/M_\mathrm{tot}$ as well as its derivative along the radial ($x\, {=} \, r$, to $501\, R_\odot$), the polar ($x\, {=} \, \cos \theta$) and the azimuthal angle ($x\, {=} \, \phi$) in (a), (b) and (c), respectively. The distributions for the total envelope and the launched tracer particle exceeding $v_r\, {=}\, 110\, \mathrm{km\, s^{-1}}$ (see Sect.~\ref{sec:jet_quants}) are plotted in violet and blue bars, respectively. The derivatives are plotted as black and gray lines. The companion is located at $r\, {=} \, 501\, R_\odot$, $\cos\theta \, {=}\, 0$ and $\phi\, {=}\, \pi$.} 
	\label{fig:tr_distribution}
\end{figure}

In Fig.~\ref{fig:tr_distribution}, we show the initial normalized radial, azimuthal, and polar mass distributions of the entire envelope, and of the mass of the tracer particles launched in the magnetic outflow (normalized by the total envelope mass and total ejected tracer particle mass, respectively). Given that the tracer particles are homogeneously distributed over the entire envelope mass for $t\,{=}\, 0 \, \mathrm{d}$ (Sect.~\ref{sec:methods}), one can directly infer how much the individual mass shells, slices and sections are contributing to the ejection through the magnetically driven outflows. Hence, if the distributions (and their logarithmic derivatives) coincide, each part of the primary star is contributing equally to the bipolar outflow along the corresponding dimension. 

Analogously, the polar (Fig.~\ref{fig:tr_distribution}b) and azimuthal (Fig.~\ref{fig:tr_distribution}c) distributions largely overlap, thus, each conical section and slice is approximately contributing equally to the magnetically driven outflows. However, there are small systematic differences. 
We find a higher contribution from the northern hemisphere as well as the upstream direction of the primary star (around $\phi \, {\approx} \, \pi/4$) and lower contribution in the downstream region (around $\phi \, {\approx} \, 3\pi/4$). The difference in the azimuthal direction can be explained due to the immediate interaction of the companion with downstream material within the first half orbit (e.g., see Movie M1 in Table~\ref{tab:movietable}, where this material is first perturbed and partially ejected).
The preference towards the northern hemisphere may be traced back to a small asymmetry in the initial position of the companion with respect to the orbital plane, and we will encounter this trend once more in Sect.~\ref{sec:tr_outflow_rates}, when we investigate the ejection rates in the outflows.

For the radial direction, however, the outer radial layers of the stellar profile (distances larger than ${\approx}\, 120 \, R_\odot$ in Fig.~\ref{fig:tr_distribution}a) are contributing less to the magnetically launched material as expected for a homogeneous distribution of tracer particles along the envelope. This is due to the outer parts of the envelope being primarily ejected during the plunge-in. However, deeper in the stellar interior, the two distributions overlap, meaning each spherical shell is evenly contributing to the magnetically driven outflow. 

This finding is rather surprising, since the companion is spiraling deeper into the initial stellar profile and one would expect an imprint of the CE evolution on the distribution of the launched tracer particles as for the outer parts of the envelope. However, the tracer particles seemingly lose memory of their original positions in the stellar profile provided they survive the plunge-in and form a disk structure.
Given that the entire envelope is ejected in our CE scenario (see Sect.~\ref{sec:mhd_vs_nomhd}), the tracer particles also serve as a proxy for the post plunge-in envelope. Consequently, the material in the circumbinary disk must also largely lose its dependence on its radial stellar origin. We attribute this finding to the turbulent enhanced transport, as discussed in \citet{vetter2024a}.

\begin{figure}
    \centering
    \includegraphics[width=\linewidth]{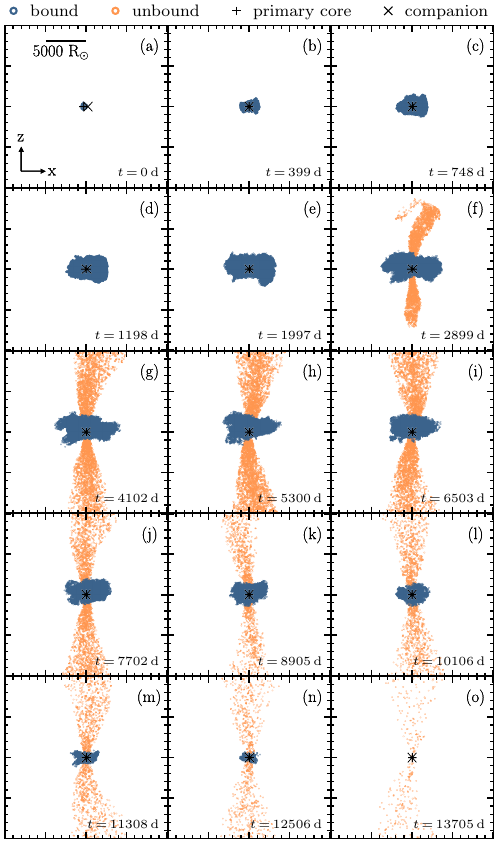}
    \caption
	{Time evolution of the position of launched tracer particles exceeding a critical velocity of $v_r\, {=} \, 110\, \mathrm{km\, s^{-1}}$ (see Sect.~\ref{sec:jet_quants}) in the $x$-$z$ plane. The colors indicate whether the particles are bound (blue) or unbound (orange) according to the kinetic energy criterion in Eq.~(\ref{eq:e_kin_crit}). The primary core and companion positions are marked with a black \qq{+} and \qq{x}, respectively.}
	\label{fig:tr_pos_timeseries}
\end{figure}

As for the trajectory of the tracer particles, during the initial rapid spiral-in of the companion (i.e., within the first $750\, \mathrm{d}$), approximately $80 \, \mathrm{\%}$ of the envelope material (and all selected tracer particles) are lifted but remain gravitationally bound to the core binary in a torus-like structure (indicated by the blue color in Fig.~\ref{fig:tr_pos_timeseries}a-\ref{fig:tr_pos_timeseries}e for $t\, {\lesssim}\, 2000\, \mathrm{d}$).
As the material stays in the orbital plane, the motion of the gas circularizes and is eventually advected towards the central binary (see \citealp{vetter2024a} and also Fig.~\ref{fig:tr_pos_timeseries}), $t\, {\gtrsim}\, 2000\, \mathrm{d}$), until the tracer particles get accelerated and ejected along the polar axis (as indicated by the orange color in Fig.~\ref{fig:tr_pos_timeseries}). Although this finding of the trajectory of the tracer particles seems trivial, it has previously not been confirmed that the magnetically driven outflow is fueled by a CBD forming from the perturbed but yet gravitationally bound envelope material as proposed in \citet{ondratschek2022a} and \citet{vetter2024a}.

\subsubsection{Acceleration region of the magnetically driven outflows}\label{sec:acce_pos}

\begin{figure}
    \includegraphics[width=\linewidth]{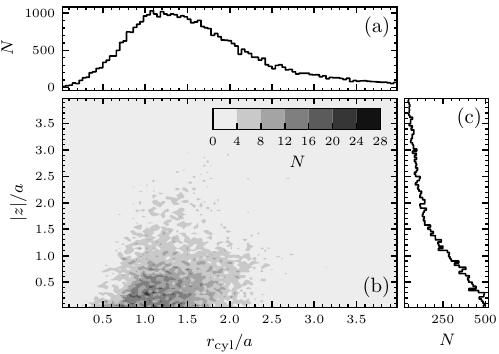}
    \caption{Corner plot of the launching position for all tracer particles exceeding a critical velocity of $v_r \,{=}\, 110\, \mathrm{km\, s^{-1}}$ (for the selection process, see Sect.~\ref{sec:jet_quants}). In (b), we show the number of tracer particles launched at a given vertical ($\vert z \vert/a$) and cylindrical radial distance ($r_\mathrm{cyl}/a$) from the center of mass of the core binary, normalized by the orbital separation at the time of acceleration. In (a) and (c), the marginalized distributions are shown. The mode of the smoothed distribution (piecewise interpolated with a cubic polynomial) in the radial direction is at $1.1a$ (i.e., the radius with the highest probability for a tracer particle to be launched).}
    \label{fig:launching_point}
\end{figure}

In the following, we try to determine the acceleration region for the magnetically driven outflows (see Sect.~\ref{sec:mag_launching}). For our analysis, we focus again on the launched tracer particles based on the least strict radial velocity criterion (i.e., $v_r\,{=} \, 110 \, \mathrm{km\,s^{-1}}$) to obtain the broadest possible overview (see Sect.~\ref{sec:jet_quants}). We perform second-order accurate time derivatives of the velocity of each tracer particle and try to identify a local extrema in the absolute value of the acceleration, where we additionally use the velocity as an identifier to acquire the correct extrema. 
In this way, we obtain the launching time, which in turn, can be translated to the position of acceleration. One must state, that the data is in general very noisy given the sparse time sampling of $3\,\mathrm{d}$). Consequently, the method does not converge for all tracer particles and about $1842$ out of $51,077$ are rejected due to difficulties. Nonetheless, we obtain a very broad overview where the particles are preferably accelerated, as we show in the corner plot in Fig.~\ref{fig:launching_point}. 
There, the distribution of the acceleration positions along the radial $r_\mathrm{cyl}=\sqrt{x^2 + y^2}$ and absolute vertical direction $\vert z\vert$, centered on the center of mass of the core binary and normalized by the orbital separation at the time of maximum acceleration, is plotted. While the vertical distribution is peaking around the mid-plane (Fig.~\ref{fig:launching_point}c), the radial distribution shows a pronounced peak around $r_\mathrm{cyl}\, {=}\, 1.1a$ (i.e., the mode of the distribution).  

\subsubsection{Measuring outflow rates}\label{sec:tr_outflow_rates}
\begin{figure}
    \centering
    \includegraphics[width=\linewidth]{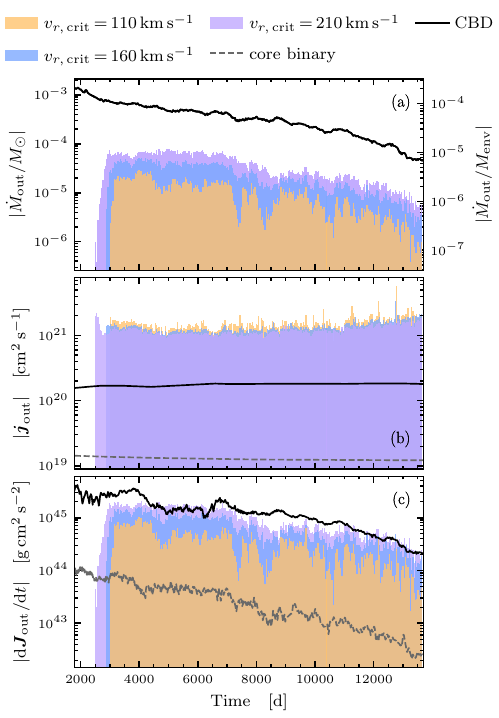}
    \caption
	{Time evolution of quantities carried away by the launched tracer particles in the polar direction. Shown are the mass-loss rate $\dot{M}_\mathrm{out}$ in (a), the specific angular momentum $\vert \boldsymbol{j}_{\mathrm{out}}\vert$ in (b) and the angular momentum loss rates $\vert \mathrm{d}\boldsymbol{J}_{\mathrm{out}}/ \mathrm{d}t\vert$ in (c). The colors represent the different critical radial velocities introduced in Sect.~\ref{sec:jet_quants} of $v_\mathrm{r,crit} \, {=}\, 110, \, 160$ and $210\, \mathrm{kms^{-1}}$ in purple, blue and orange, respectively. We additionally show the respective quantities for the CBD in black and for the core binary in dashed gray. With the subscript \qq{out}, we specifically highlight, that the rates shown in this plot are obtained through loss rates of tracer particles passing a control surface within a time window of $42\, \mathrm{d}$ (see Sect.~\ref{sec:jet_quants} for more details).}
	\label{fig:outflow_binned_BH}
\end{figure}
In order to measure the properties of the magnetically driven outflow, we count the previously selected tracer particles crossing through a spherical surface with radius $r \, {=} \, 5000 \,R_\odot$ centered on the center of mass of the core binary within time intervals of $\Delta t \, {=} \, 42\, \mathrm{d}$.
This way, we get the time evolution of the mass outflow rate $\dot{M}_\mathrm{out}$, total specific angular momentum $\vert \boldsymbol{j}_\mathrm{out}\vert $ and the angular momentum loss rate $\vert \mathrm{d}\boldsymbol{J}_\mathrm{out}/\mathrm{d}t \vert $ through the bipolar outflow (Fig.~\ref{fig:outflow_binned_BH}). Our findings throughout the different radial velocity criteria are summarized in Table~\ref{tab:quant_jet_out} and, in the following, we will again illustrate our results for the least strict criterion (i.e., $v_{r, \, \mathrm{crit}}\,  {=} \, 110 \,\mathrm{km \, s^{-1}}$, see Sect.~\ref{sec:jet_quants}) to obtain the broadest possible overview. 

In total, the tracer particles carry away a mass of $M_\mathrm{out,\, tot}\, {=}\, 0.42 \, M_\odot \, {\approx}\, 0.064 \, M_\mathrm{env}$ and the total launched angular momentum amounts to $J_\mathrm{out,\, tot}\, {\approx} \, 9.5 \, {\times} 10^{53} \, \mathrm{g\, cm^2\,s^{-1}}$.
The mass-loss rate through the bipolar outflow is approximately an order of magnitude lower than the overall mass ejection rate (Fig.~\ref{fig:outflow_binned_BH}a).

We find an average mass-loss rate of $\langle \dot{M}_\mathrm{out}\rangle_\mathrm{t}\, {=}\, 3.1 \, {\times}\, 10^{-5}\,M_\odot\,\mathrm{d^{-1}}$ (see Table~\ref{tab:quant_jet_out}, with $\langle Q \rangle_\mathrm{t} \, {\equiv} \, \sum_j Q_j/\sum_j 1$, the arithmetic mean of quantity $Q$ over all time bins $j$) which is roughly evenly distributed among both polar directions with a slight preference towards the northern hemisphere ($\langle \chi_\mathrm{sym} \rangle \, {\approx} \, 1.2 \, \%$  more mass is ejected through the upper funnel). A similar preference towards the northern hemisphere is already observed in the initial distribution of the tracer particles that are launched in the bipolar outflows (see Sect.~\ref{sec:tr_origin}) and is caused by slight offset of the companion in the initial condition.
Given that the magnetized outflow is refueled by the CBD (Sect.~\ref{sec:tr_origin}) and in confirmation of the statements in \citet{vetter2024a}, these rates recover the reported inward mass fluxes from the CBD material and are roughly a factor of ten smaller than the recombination-driven winds (see Fig.~\ref{fig:outflow_binned_BH}a). 
However, they largely exceed the typical mass-loss rates found in PN observations for jets in post-asymptotic giant branch (AGB) binary systems \citep[e.g.][]{bollen2020a,bollen2022a}, 
or simulations \citep[e.g.,][]{garcia2021a,ondratschek2022a} by approximately one to four orders of magnitude. The mass-loss rates are about ${\sim}\, 10^{-7}\,\textrm{--}\,10^{-3} \,{\times}\, \mathrm{M_\odot\, yr^{-1}}\, {=}\, 10^{-9}\,\textrm{--}\,10^{-6} \,{\times}\, \mathrm{M_\odot\,d^{-1}}$ in these CE systems with considerably lower-mass AGB primary stars. 

Furthermore, we find that the angular momentum loss rates of the magnetically driven outflows almost entirely balance the ones of the CBD, while the loss rates of the core binary remain about one order of magnitude lower. The average angular momentum loss rates are $\langle \mathrm{d}\boldsymbol{J}_\mathrm{out}/\mathrm{d}t \rangle_\mathrm{t} \,{\approx}\, 8.05\, {\times} \,10^{44}\, \mathrm{g\, cm^2\,s^{-2}}$.
The specific angular momentum extracted by the outflows ($\vert \boldsymbol{j}_\mathrm{out}\vert$) remains relatively constant over time (see Fig.~\ref{fig:outflow_binned_BH}b) with average values of $\langle \boldsymbol{j}_\mathrm{out} \rangle_\mathrm{t}\, {\approx}\, 1.2\, {\times}\, 10^{21}\, \mathrm{cm^2 \, s^{-1}}$. 
On average, the bipolar outflows extract $\langle j_\mathrm{out}/j_\mathrm{K}(r\,{=}\,1.1a)\rangle_\mathrm{t}\, {=}\, 21.4$ times the Keplerian specific angular momentum at the launching point (i.e., $j_\mathrm{K}(r\,{=}\,1.1a)\, {=} \, \sqrt{GM_\mathrm{b}1.1a}$, see Sect.~\ref{sec:tr_origin} and Sect.~\ref{sec:acce_pos}). In fact, this fraction is roughly consistent with the expectations for a Blandford--Payne like outflow: the fraction of ejected and Keplerian specific angular momentum at the launching point of the disk ($j_\mathrm{out}$ and $j_\mathrm{K}$) is equal to the squared fraction of the distance to the Alfvén surface ($R_\mathrm{A}$) and the distance to the launching point (labeled as $R$ here), $j_\mathrm{out}/j_\mathrm{K}(R) \, {=}\, (R_\mathrm{A}/R)^2$. For instance, from Fig.~\ref{fig:magn_launching}, one can infer that $R_\mathrm{A}\, {\approx} \, 200 R_\odot$, which yields values ranging from $j_\mathrm{w}/j_\mathrm{launch} \, {=}\, 19\, \textit{--}\, 35$ for an orbital separation of $37\, \textit{--}\, 50 \, R_\odot$.  

The number-weighted average velocity of the tracer particles in the outflow is $\langle v_r\rangle_\mathrm{p} \, {=}\,  172 \,\mathrm{km\,s^{-1}}$ (here, $\langle Q \rangle_\mathrm{p} \, {\equiv} \, \sum_i Q_i/\sum_i 1$, where the summation is taken over each tracer particle $i$ for a given selection method). The time averaged velocity, $\langle v_r\rangle_\mathrm{t} \, {=}\,  172 \,\mathrm{km\,s^{-1}}$, shows only minor variations in the evolution, with a standard deviation of ${\approx}\, 14 \, \mathrm{km \, s^{-1}}$. This further supports the conclusion that the magnetically driven acceleration mechanism create continuous outflows. 
Comparing the radial outflow velocity to the orbital velocity of the core binary, we find typical values of $\langle v_r/v_\mathrm{orb}\rangle_\mathrm{t} \, {=}\, 0.89$ averaged over time and $0.91$ per tracer particle. Moreover, the average kinetic energy carried away by the outflow (i.e., $0.5M_\mathrm{out}\langle v_r\rangle_\mathrm{p}^2 \, {\approx}\, 1.3\, {\times} \, 10^{47} \, \mathrm{erg}$) matches the difference in gravitational energy between the Hydro and MHD simulation (i.e., $0.5GM_2M_\mathrm{c}(a_{f,\, \mathrm{MHD}}^{-1} - a_{f,\, \mathrm{Hydro}}^{-1})\, {\approx} \, 1.5\, {\times}\, 10^{47}\, \mathrm{erg}$). 

A comparison between the quantities throughout the regions of the magnetized outflows spanned by the critical velocities can be found in Table~\ref{tab:quant_jet_out}. For higher critical velocities, the overall picture set forth above remains, but the individual quantities can vary, yet, they remain on the same order of magnitude. 

\begin{table}
\centering
\caption{Properties of tracer particles ejected through the bipolar outflows (see Sect.~\ref{sec:jet_quants}).}
	\label{tab:quant_jet_out}
    \begin{tabular}{l|ccc}
	\hline\hline
    Quantity & \multicolumn{3}{c}{Values} \\ 
	$v_{r, \mathrm{crit}} \quad \mathrm{[km \, s^{-1}]}$                                               &110&160&210 \\
	\hline
	$\langle v_r\rangle_\mathrm{p}\quad \mathrm{[km \, s^{-1}]}$                                       &174	&201 &228\\
	$\langle v_r\rangle_\mathrm{t}\quad \mathrm{[km \, s^{-1}]}$                                       &172	&199 &225\\
    $\langle v_r/v_\mathrm{orb}\rangle_\mathrm{t}$                                                     &0.89 &1.02 &1.15\\
    $\langle v_r/v_\mathrm{orb}\rangle_\mathrm{p}$                                                     &0.91 &1.04 &1.18\\ 
	$N_\mathrm{out, \,tot}$	                                                                           &51,077	&30,811&12,649 \\
	$M_\mathrm{out, \, tot} \quad [M_\odot]$                                                           &0.42&0.25&0.11 \\
	$M_\mathrm{out, \, tot} \quad [M_\mathrm{env}]$                                                    &0.064&0.039&0.016 \\
    $\langle M_\mathrm{out}\rangle_\mathrm{t} \quad [10^{-3}\, {\times}\, M_\odot]$                    &1.31&0.79&0.32 \\
    $\langle \dot{M}_\mathrm{out}\rangle_\mathrm{t}\quad \mathrm{[10^{-5}\, {\times}\, M_\odot \, d^{-1}]}$	   &3.1	&1.9&0.8\\
    $\langle| \boldsymbol{j}_\mathrm{out} |\rangle_\mathrm{t} \quad \mathrm{[10^{21}\, {\times}\, cm^2 \, s^{-1}]}$    &1.21&1.36&1.58\\
    $\langle |\boldsymbol{j}_\mathrm{out}| /j_\mathrm{K}(r\,{=}1.1a)\rangle_\mathrm{t}$                   &21.4&24.0&27.9\\
	$\langle |\mathrm{d} \boldsymbol{J}_\mathrm{out} /\mathrm{d}t|\rangle_\mathrm{t}\quad [\mathrm{10^{44}\, {\times}\, g\, cm^2\, s^{-2}}]$  &8.05&5.46&2.59\\
 	$\langle \chi_\mathrm{sym}\rangle_\mathrm{t}\quad [\mathrm{\%}]$		                                       &1.2	&1.3 &7.9\\
	\hline
	\end{tabular}	
    \tablefoot{This table shows the averaged radial velocity $\langle v_r\rangle$, the radial velocity normalized by the orbital velocity of the core binary $\langle v_r/v_\mathrm{orb}\rangle$, the total number of launched tracer particles $N_\mathrm{out, \, tot}$, the total ejected mass through the magnetic outflow $M_\mathrm{out,\, tot}$, the average ejected mass $\langle M_\mathrm{out}\rangle$, the averaged mass-loss rate $\langle \dot{M}_\mathrm{out}\rangle$, the averaged specific angular momentum loss $\langle | \boldsymbol{j}_\mathrm{out}| \rangle$, the loss normalized by the Keplerian value at the launching position $\langle j_\mathrm{out} / j_\mathrm{K}(r\, {=}\, 1.1a) \rangle$ (see Sect.~\ref{sec:acce_pos} for the launching point and $j(r\, {=}\, 1.1a) \, {=}\, \sqrt{GM_\mathrm{b} 1.1 a}$ with $M_\mathrm{b}$ and $a$ being the core binary mass and distance), the average angular momentum loss rates $\langle | \mathrm{d} \boldsymbol{J}_\mathrm{out} /\mathrm{d}t |\rangle$ and the averaged symmetry measure ($\chi_\mathrm{sym} \, {=}\, N_\mathrm{out, \, north}/N_\mathrm{out, \, tot}\,{-} \,0.5$, with $N_\mathrm{out, \, north}$ is the amount of tracer particles ejected through the northern hemisphere) for each individual critical radial velocity $v_{r, \mathrm{cit}}$ defined in Sect.~\ref{sec:jet_quants}. The averages index with \qq{$\mathrm{p}$} are averages per tracer particle with $\langle Q \rangle_\mathrm{p} \, {\equiv} \, \sum_i Q_{i}/\sum_i 1 $ (where $i$ runs over all tracer particles selected by a particular critical velocity), while $\langle Q \rangle_\mathrm{t} \, {\equiv} \, \sum_\mathrm{j} Q_\mathrm{j}/\sum_\mathrm{j} 1$ are averages over time bins with index $j$ in Fig.~\ref{fig:outflow_binned_BH} (hence, representing averages over time).}
\end{table}

\subsection{The neutron star companion}\label{sec:ns_comp}

As stated in \citet{vetter2024a}, we observe magnetically driven outflows in both CE simulations involving a BH and a NS companion. While most findings in Sect.~\ref{sec:mag_ampl} and Sect.~\ref{sec:mag_launching} apply in both cases, we also observe differences between the models. 
To quantify these differences, we apply the same analysis of the tracer particles to the model involving the NS (see Appendix~\ref{sec:appendixB}, Table~\ref{tab:quant_jet_out_NS}). 
As already stated in \citet{vetter2024a}, we observe significant rotation of the orbital plane of the core binary with respect to its initial orientation in the NS scenario, 
which leads to interactions of the launched and the CBD material.  
Therefore, unlike in the case of the BH companion, the properties of the jetted outflow may be altered by its interaction with CBD material along its launch direction. 

We find similar total specific angular momentum and mass-loss rates in the outflows (also for the core binary and the CBD), although both are subject to larger fluctuations, most likely caused by the pronounced evolution of the orbital plane's orientation \citep[][]{vetter2024a}. Similar to the $q\, {=}\, 0.5$ case, the angular momentum loss rate of the core binary is two orders of magnitude below the rates found for the outflows and bound material, while the rates approximately recover the ones in the CBD. The specific angular momentum lost within the time bins is approximately constant and matches the results found in the BH companion scenario of about $\langle | \boldsymbol{j}_\mathrm{out}|\rangle_t \, {\approx}\, 1.16\, {\times}\, 10^{21} \, \mathrm{cm^2\,s^{-2}}$.

We observe that the post plunge-in orbital contraction rates are in both scenarios similar, which is consistent with our findings in Figs.~\ref{fig:outflow_binned_BH}, \ref{fig:outflow_binned_NS} and Tabular.~\ref{tab:quant_jet_out}, \ref{tab:quant_jet_out_NS} regarding the angular momentum extraction from the core binary.  

We find slightly higher radial velocities for a higher mass ratio. Hence, the outflows reach radial velocities up to $263\, \mathrm{km\,s^{-1}}$ and on average $164\, \mathrm{km\,s^{-1}}$, although the orbital velocity of the core binary is found to be higher (at the end of the simulation, $v_\mathrm{orb} \, {=}\, 230 \, \mathrm{km \, s^{-1}}$) compared to the BH companion scenario. Consequently, the average of the fraction of radial velocity and orbital velocity of the core binary is reduced from $0.89$ to $0.69$.  

This finding is opposite to the conclusion in \citet{ondratschek2022a}, where the outflow velocities are stronger for less massive companions, given the increased orbital velocity in the tighter orbits of the core binary. It is likely that factors other than the orbital velocity affect the outflow velocity, and wider parameter space exploration could help achieve a complete understanding.

\section{Discussion}\label{sec:discussion}
While this publication focuses on magnetic fields and their impact on the CE evolution, our preceding publications provide extensive discussion on the implications of our initial model and numerical setup on the CE evolution \refbold{(especially on the orbital separation, eccentricity of the core binary, and envelope ejection)}, the CBD structure and the final fate of the binary \citep{moreno2022a, vetter2024a, wei2024a}.
In the following discussion, we focus on the impact of magnetic fields on the CE phase (Sect.~\ref{sec:dis_impact_CE}), magnetic-field amplification and the launching of the magnetically driven outflows (Sect.~\ref{sec:dis_ampl_launching}) and lastly, how one can implement our findings in analytical recipes and models that do not explicitly follow the evolution of magnetic fields (Sect.~\ref{sec:dis_tension}).

\subsection{Magnetic-field amplification and launching of bipolar outflows}\label{sec:dis_ampl_launching}
In Sections \ref{sec:mag_ampl} and \ref{sec:mag_launching}, we analyzed the amplification of magnetic fields and the subsequent launching of material perpendicular to the orbital plane of the central binary.
Regarding the first phase of the amplification of magnetic fields, we identified similar mechanisms to those observed in lower-mass systems. 
However, it is essential to note that other mechanisms might play a significant role, such as, for example, the $\alpha\Omega$-mechanism \citep[e.g.,][]{kiuchi2024a, pakmor2024a} or other turbulent dynamos. Our analysis is not exhaustive, necessitating more detailed investigations to fully comprehend the amplification processes at play.
Special attention must be given to the first amplification phase, because this phase depends sensitively on initial conditions and resolution, which, as argued in \citet{gagnier2024b}, leads to underestimated shear rates in numerical simulations. 
For instance, we neglect any precursory fields existing in the convective envelope of the red supergiant or inherited by the envelope through the preceding phase of binary mass transfer. We also do not account for magnetic fields from the companions themselves (e.g., magnetized white dwarfs, NSs or BHs). Furthermore, we do not resolve the physics nor the scales of actual compact companions on our grid. Moreover, our mass refinement approach is not well-suited to accurately capture magnetic-field amplification in low-density regions, leading to underestimated shear rates as argued by \citet{gagnier2024b}, although the fastest growing mode of the MRI appears to be resolved in our case. Hence, our results can only be seen as lower limits on the actual growth time in Phase~1 of the magnetic energy. However, as argued by \citet{ohlmann2016b}, the saturation level in Fig.~\ref{fig:mag_ampli} after Phase~2 should be largely independent of the resolution and initial seed field.

Regarding the launching of material, we have identified three mechanisms which contribute to the overall acceleration of material in the polar direction. In agreement with \citet{ondratschek2022a}, we find evidence for the Blandford--Payne and the tower jet mechanisms, which are responsible for creating a steady and collimated bipolar outflow with a wound-up magnetic-field structure (see Sect.~\ref{sec:mag_launching} and Fig.~\ref{fig:overview_bfield}). Indeed, the launching of material is associated with enhanced Maxwell stresses dominating over centrifugal forces and magnetic pressure gradients, which points towards a magneto-centrifugally driven acceleration near the core binary (Sect.~\ref{sec:mag_launching}, Fig.~\ref{fig:magn_launching}). 
In addition \citep[and in agreement with][]{gagnier2024b}, we identified shock-heating of gas close to the central binary to be efficient in accelerating material into the cleared funnel in the polar direction in the Hydro case. This mechanism may also contribute to the overall acceleration in the MHD case, but can not drive the observed strong outflows in our case. Similar to the arguments presented in \citet{huggins2012a,blackman2014a,zou2020a}, our results suggest, that the outflows from PNe cannot be explained by the purely hydrodynamical acceleration mechanism.

\subsection{Impact of magnetic fields on the CE event}\label{sec:dis_impact_CE}

In contrast to \citet{ondratschek2022a}, we observe differences in orbital separation and envelope ejection between the strictly hydrodynamical and magnetized models in the CE evolution of our higher-mass system. As stated in Sect.~\ref{sec:mhd_vs_nomhd}, the simulation develops bipolar outflows that extract angular momentum from the CBD and the core binary. Consequently, the binary hardens more compared to the purely hydrodynamical simulation. Assuming circular orbits of the core binary and no mass transfer or accretion, the angular momentum of the binary is $J_\mathrm{b} \, {=}\, \mu \sqrt{GM_\mathrm{b}a}$, where $\mu$ is the reduced mass of the core binary. From that, we can estimate the specific torque difference between the Hydro and the MHD simulation $(\dot{J}_\mathrm{MHD} - \dot{J}_\mathrm{Hydro})/M_\mathrm{b}$, which reaches a minimum of $-2.5\,{\times}\ 10^{9} \, \mathrm{cm^2\, s^{-2}}$ at about $3100\,\mathrm{d}$ and increases linearly to zero until the end of the simulation. This difference is also reflected in the angular momentum loss rate from the core binary in Fig.~\ref{fig:outflow_binned_BH}, given that the core binary in the Hydro run hardly contracts further after $3500\, \mathrm{d}$ (see Fig.~\ref{fig:MHD_vs_noMHD_oe_fej}) and the difference is thus dominated by the MHD simulation. The increased stresses on the core binary system may indirectly originate from turbulent enhanced transport rates in the magnetically saturated CBD material by an additional Maxwellian contribution (${\propto}\, B_r B_\phi$) \citep[see][]{vetter2024a}. However, to fully understand the angular momentum transport in our simulations, further and more thorough investigations are required, as, for instance, is done in \citet[][]{gagnier2023a,gagnier2024a}.
\refbold{Furthermore, in the late stages of the evolution ($t\, {\gtrsim} \, 3000 \, \mathrm{d}$), we observe changes in eccentricity in the MHD simulation compared to the Hydro scenario (see Fig.~\ref{fig:MHD_vs_noMHD_oe_fej}). As discussed in more detail in \citet{moreno2022a} and \citet{vetter2024a}, the eccentricity is numerically not converged in our simulations, but a physical origin can also not be excluded. Future work is warranted to better understand the final eccentricity in our models.}

While a more detailed analysis is required to fully unravel the implication of turbulent transport on the CE interactions, we can already draw some conclusions from the results presented in \citet{ondratschek2022a} and this work. The fact, that the outflows appear to form self-consistently in models with different parameters indicates ubiquity of this phenomenon in CE events. However, we expect limits to this universality. For instance, for lower $M_2$, the core and the secondary object can merge or only mildly perturb the envelope\footnote{For example, in the planetary engulfment scenario \citep[e.g.,][]{lau2025a}, the released orbital energy is orders of magnitude smaller than the binding energy of the star and one may not expect outflows as seen in our simulations.} and neither the magnetic fields can be amplified sufficiently nor the remaining envelope can form a disk-like structure. 

In the following discussion, we assume the formation of bipolar outflows and a CBD. The outcome of the CE phase then boils down to the transport of angular momentum and the interaction of the core binary with the disk structure. In the light of enhanced turbulent transport in an effective $\alpha$-viscous model \citep[][]{shakura1973a}, one expects $\alpha$ values in the range of $10^{-2}$ to $10^{-1}$ in the saturated state throughout models \citep[e.g.][]{davis2010a}, but the transport rate will also largely depend on the disk properties such as aspect ratio, local densities and sound speed. While the aspect ratio might arguably be insensitive through different CE scenarios (a thick disk with aspect ratios close to unity seems to form post plunge-in), the local densities and sound speeds in the CBD will depend on the initial primary star and companion model.
For the duration of the post plunge-in interaction of the core binary with the CBD \citep[see the systems in, e.g.,][]{vetter2024a}, the time increases for increasing $M_2$ and extends even beyond the dynamical interaction as the CE event becomes less adiabatic \citep[e.g.,][]{podsiadlowski2001a}. As a result, a long-lived co-existence of the core binary and a disk can be established \citep[e.g.,][]{tuna2023a,wei2024a,vetter2024a}.
In light of these thoughts, it is even more surprising that for different companion masses and the given primary model, the relative change in final orbital separation between the models with and without magnetic fields should be comparable, as we argued in Sect.~\ref{sec:ns_comp}, and future investigations are required to understand this.

For the envelope ejection process, we do not see significant differences in our simulation between the Hydro and MHD case. The first mass-loss episode can be traced back to the dynamical ejection of about $18\, \textit{--}\, 20\, \mathrm{\%}$ of the envelope mass (for \refbold{$t\, {\lesssim} \, 1500\, \mathrm{d}$}) due to the plunge-in of the companion (see Fig.~\ref{fig:MHD_vs_noMHD_oe_fej} and also Appendix~\ref{sec:appendixA}, Fig.~\ref{fig:ekin_evolution} for the kinetic energy density evolution), and the MHD and the Hydro simulations coincide. For the subsequent mass ejection event, recombination-driven winds at the edge of the CBD take over and strip the remaining envelope, with mass-loss rates on the order of ${\sim}\, 10^{-3}\, \textit{--}\, 10^{-5}\, M_\odot\,\mathrm{d}^{-1}$. These rates exceed those found in the bipolar outflows by about an order of magnitude and, hence, recombination-driven winds largely shape the overall envelope ejection evolution. 
\refbold{The recombination-driven winds are likely sensitive to radiative cooling in the remaining envelope material, but this is not included in our adiabatic simulations \citep[see the discussion in][]{vetter2024a}. As argued in \citep[e.g.,][]{meyer1979a,clayton2017a,fragos2019a,bronner2023a} and recently in \citet{lau2025b}, the extended bound envelope can exhibit short radiative transport timescales in the late stages of the interaction and may cause the dominant equatorial envelope ejection to cease.}

\subsection{Incorporating magnetically driven outflows in other common-envelope models}\label{sec:dis_tension}

Based on the results presented here and in \citet{vetter2024a}, we develop effective models that allow for the integration of our main results in simplified models of common-envelope evolution. We focus on two aspects: firstly, the formation of bipolar outflows, impacting the orbital separation and envelope ejection already in the dynamical phase of the evolution and, secondly, the implementation of angular momentum transport and extraction from a disk-like structure in the post-plunge-in system. 

For the first point, we can adjust the $\alpha_\mathrm{CE}$-energy formalism \citep{heuvel1976a, webbink1984a} such that we account for the kinetic energy lost through the bipolar outflows in the equation (as argued in Sect.~\ref{sec:tr_outflow_rates}). Since the average ejection velocity per particle is on the order of the orbital velocity of the core binary (see Table~\ref{tab:quant_jet_out}), we can approximate the kinetic energy carried away by the outflows $0.5M_\mathrm{out}\langle v_r\rangle_\mathrm{p}^2\,{\approx}\, 0.5 M_\mathrm{out} GM_\mathrm{b}/a_f$ with $M_\mathrm{out}$ (Sect.~\ref{sec:tr_outflow_rates}), which leads to a modification of the energy formalism:
\begin{align}
    \alpha_\mathrm{CE}\left[ -\frac{GM_\mathrm{c}M_2}{2a_f}\left (1+\frac{M_\mathrm{out}}{\mu} \right ) + \frac{GM_1M_2}{2a_i}\right] = E_\mathrm{bind}, \label{eq:alpha_CE}    
\end{align}
where $E_\mathrm{bind}$ is the binding energy of the envelope. This implies that the relative change in final orbital separation for the MHD versus Hydro case is proportional to the fraction of the mass lost in the wind divided by the reduced mass of the core binary $a_{f,\, \mathrm{Hydro}}/a_{f,\, \mathrm{MHD}} \, {=}\, 1 + M_\mathrm{out}/\mu$, which we argued, must be approximately the same for the NS and the BH case due to the insensitivity of the angular momentum loss rates to $M_2$ (Sect.~\ref{sec:ns_comp}). In fact, we find $M_\mathrm{out}/\mu$ to be largely independent of the secondary star mass (for $v_{r, \, \mathrm{crit}}\, {=}\, 110,\,160,\, 210\, \mathrm{km\, s^{-1}}$, we find $0.22,\, 0.13,\, 0.06$ and $0.19,\, 0.12,\, 0.04$ for the BH and NS companion, respectively). Furthermore, we can express the mass ejected through the outflow as the time integral over the mass-loss rate $M_\mathrm{out}\, {=}\, \int_{\Delta t} \dot{M} \, \mathrm{d}t$, where we integrate from the onset of the bipolar outflow to the end of the simulation. As stated in Sect.~\ref{sec:jet_quants}, the mass-loss rates through the outflows recover the radially inwards mass transport rates in the CBD, which we can express in terms of an effective turbulent $\alpha$-viscosity \citep[see][]{vetter2024a}. Hence, the ejected mass can be approximated as $M_\mathrm{out}\, {=}\, 3\pi \alpha \theta^2 \Omega \int_{\Delta t} \Sigma(R,t)\, \mathrm{d}t$ with $\Omega$ and $\theta$ being the angular velocity and aspect ratio (which are mostly independent of time) and $\Sigma$ the integrated column density of the CBD at radius $R$. As one can already infer from this expression, the quantities involved are expected to differ for different primary star models and, for instance, we find $M_\mathrm{out}/\mu\, {\approx} \, 0.024$ for the mass ejection rates reported in \citet{ondratschek2022a} in their $q\,{=}\,0.25$ scenario (assuming a linear decay of $\dot{M}$ in time and integrating to full envelope ejection from about $1500\, \mathrm{d}$ to $4500\, \mathrm{d}$ leads to $M_\mathrm{out}\, {\approx}\, 0.004 \, M_\odot$). As a consistency check, this translates into a final orbital separation of $21.5\,R_\odot$ in their Hydro scenario, coinciding perfectly with the reported final orbital separation of $21.5\, R_\odot$ (taking the reported final orbital separation of $21.0\, R_\odot$ for their MHD simulation).
Interestingly, the values for $M_\mathrm{out}/\mu$ in this work and the ones for \citet{ondratschek2022a} differ by a factor of ten, which is reflected in the quotient of the initial primary star masses.

As for our second point regarding the extraction of angular momentum from the CBD, one can include an effective $\alpha$-viscosity model to account for the enhanced angular momentum transport found in our MHD case in non-magnetized one-dimensional (1D) or multidimensional treatments of the CE interaction. Moreover, an ad hoc central engine with the parameters found in Sect.~\ref{sec:jet_quants} can be employed as used, e.g., in \citet{grichener2021a} and \citet{dori2023a} for 1D models and, e.g., in \citet{garcia2018a}, \citet{zou2020a}, \citet{garcia2020a}, and \citet{garcia2021a} for multidimensional models. 

\section{Conclusions}\label{sec:conclusion}
In this work, we investigated the magnetic-field amplification and the launching of magnetized outflows, as well as their implications for the CE evolution of a $10\, M_\odot$ RSG and a $5\, M_\odot$ BH companion. Our results can be summarized as follows:
\begin{itemize}
    \item In contrast to the findings in \citet{ondratschek2022a}, we do observe considerable differences in the orbital separation and envelope ejection between the purely hydrodynamical and magnetized CE evolution. 
    The final orbital separation of the MHD model is decreased by roughly $24\, \mathrm{\%}$ from $46\, R_\odot$ to $37\, R_\odot$. This difference can be traced back to the enhanced angular momentum transport and the presence of magnetically driven outflows. Consequently, angular momentum is extracted from the core binary by increased torques (as argued in Sect.~\ref{sec:dis_impact_CE}). 
    For the envelope ejection, we only observe temporary differences in the ejection process. The deviations are presumably caused by the difference in angular momentum transport; however, the recombination-driven equatorial winds dominate the ejection in our adiabatic setup and the differences disappear by the end of the simulation.
    
    \item We further observe an increase of the magnetic energy of about $10$ orders of magnitudes, where we subdivided the evolution into four distinct phases (see Sect.~\ref{sec:mag_ampl}). The first phase \citep[rapid amplification phase,][]{ohlmann2016a} takes place well within the first orbit and is characterized by strong shear rates during the plunge-in. The second phase (slow amplification phase) is initiated as the spiral arm, created within the first orbit, becomes unstable and amplifies the magnetic field in the wake and close to the shock front. The second phase is followed by a decrease of the magnetic-field strength (Phase 3) due to magnetic flux conservation and the expanding envelope. 
    And last, the fourth phase, where we again observe strong shear -- this time near the central binary -- and simultaneously the onset of a bipolar magnetically driven outflow (Sect.~\ref{sec:mag_launching}).
    Our findings are in agreement with earlier works for different CE models \citep{ohlmann2016a, ondratschek2022a}. The fact that the evolution is similar in distinct high- and low-mass systems points to a universal amplification mechanism in CE evolution. Establishing this universality firmly, however, requires a wider exploration of the parameter space.
    
    \item While we also observe weak intermittent bipolar ejection in the purely hydrodynamical simulation \citep[similar to][]{gagnier2024b}, the MHD simulation produces strong, stable and collimated ejection of material and angular momentum in the polar direction. In agreement with simulations of low-mass systems \citep{ondratschek2022a} and even double white dwarf and white dwarf-NS mergers \citep[e.g.,][]{moran2024a, pakmor2024a}, we see indications for both the Blandford--Payne mechanism \citep{blandford1982a} and tower-jet-like acceleration. Also in agreement with \citet{garcia2021a} and their wide-jet prescription, we find increased Maxwell stresses and a steep magnetic pressure gradient in the launching zone. As expected for a magneto-centrifugally launched outflow from a disk, we observe a helical magnetic-field structure induced by the circular motion of the gas in the bound material. The material is accelerated up to about $v_r \, {\approx}\, 290 \, \mathrm{km \, s^{-1}}$ (on average to $174 \, \mathrm{km \, s^{-1}}$). \refbold{The structure that arising shares morphological similarities with bipolar PNe \citep[see][]{ondratschek2022a,vetter2024a} and magnetized post-AGB systems with maser emissions \citep[so called water fountains, e.g.,][]{vlemmings2006a,khouri2021a,khouri2025a}} 
    
    \item We further analyze the outflow with the help of tracer particles (Sect.~\ref{sec:jet_quants}). We find that the accelerated material is in large parts homogeneously distributed over the initial stellar profile, except for the outer parts of the envelope $r\, {\gtrsim} \, 120\,R_\odot$, which is primarily ejected during the plunge-in. We attributed this finding to efficient mixing in the CBD post-plunge-in, which is induced by enhanced turbulent transport \citep[see Sect.~\ref{sec:tr_origin}, Sect.~\ref{sec:dis_impact_CE} and further][]{vetter2024a}. 
    For $t\, {\gtrsim}\, 2000\,\mathrm{d}$, the tracer particles are progressively ejected in the polar directions, where the acceleration area is located around $1.1$ times the orbital distance of the core binary and tightly scattered around the orbital plane (see Sect.~\ref{sec:acce_pos}).
    Furthermore, we constrained the transport rates of the bipolar outflow (see Sect.~\ref{sec:jet_quants}), where we find that ${\approx} \, 6.4 \, \mathrm{\%}$ of the envelope mass (nominally $0.42 \, M_\odot$) are ejected with an average mass-loss rate of $3.1\,{\times} \, 10^{-5}\, M_\odot \, \mathrm{d}^{-1}$. These rates are in agreement with the inward mass transport rates in the CBD \citep[cf.,][]{vetter2024a} and about one order of magnitude below the loss rates through recombination-driven winds. 
    The angular momentum extraction by the outflow is consistent with a Blandford--Payne-like outflow (extracting $21.4$ times the Keplerian specific angular momentum at the launching point) and on average we find angular momentum loss rates of $8.05\, {\times}\, 10^{44}\,\mathrm{g^{-1}\, cm^{2}\,s^{-2}}$ balancing the rates in the CBD and exceeding the the ones for the core binary by about one order of magnitude (see Fig.~\ref{fig:outflow_binned_BH} and Table~\ref{tab:quant_jet_out} for more details).
    From the trajectories of the tracer particles (Sect.~\ref{sec:tr_origin}), we concluded that the outflows are refueled by the remaining bound CBD material.
    
    \item As for the neutron star companion, the magnetic-field amplification as well as the launching mechanisms are identical to those observed in Sects.~\ref{sec:mag_ampl} and~\ref{sec:mag_launching}, and similar to the findings in \citet{ohlmann2016b} and \citet{ondratschek2022a}. Compared to the $q\, {=} \,0.5$ system, the specific angular momentum ejection and mass-loss rates are found to be comparable to the BH companion case, while the total ejected mass is reduced to $0.18\, M_\odot$ (or $2.7\, \mathrm{\%}$ of the envelope). From the similarities between the two simulations, we concluded that the orbital contraction rates induced by the outflows must be comparable in both scenarios.
    
    \item Based on our findings in this and previous work \citep{ondratschek2022a,vetter2024a}, we discuss the potential implementation of the effects of a magnetized CE interaction in analytical and numerical approaches. We find that the final orbital energy in the $\alpha_\mathrm{CE}$-formalism can be adapted by introducing a factor of $1+M_\mathrm{out}/\mu$ and numerical approaches may include a turbulent $\alpha$-model (\citealp{shakura1973a}, for which we reported an effective $\alpha$ of $0.06\,\textit{--}\,0.13$ in \citealp{vetter2024a}) in combination with an ad hoc central engine based on the results reported in Table~\ref{tab:quant_jet_out} and \ref{tab:quant_jet_out_NS}, to mimic the angular momentum loss by the magnetically driven outflows.       
\end{itemize}

In conclusion, we find that the involvement of magnetic fields in the simulation can not only affect the orbital separation but also lead to the self-consistent formation of magnetically driven bipolar outflows, which appears to be a robust outcome of the dynamic interaction.
It must be noted that our findings and analysis are far from being exhaustive.
Future investigations may look into the acceleration area, conduct a more detailed analysis of the field amplification (especially Phase 4), shed light on the mechanisms behind angular momentum transport, refine our understanding of the mechanism behind the vertical acceleration in the bipolar outflows (potentially with tracer particles), and study the sensitivity of our findings with regard to the mass ratio and mass of the primary star. 

\begin{acknowledgements}
    \refbold{We thank the referee, Dr.~Zhuo Chen, for the constructive and helpful review process, which improved the quality of the article.}. M.V., F.K.R., F.R.N.S., R.A., M.Y.M.L and D.G.\ acknowledge support by the Klaus-Tschira Foundation. M.Y.M.L. has been supported by a Croucher Foundation Fellowship. M.V., F.K.R., R.A. and D.G.\ acknowledge funding by the European Union (ERC, ExCEED, project number 101096243). Views and opinions expressed are, however, those of the authors only and do not necessarily reflect those of the European Union or the European Research Council Executive Agency. Neither the European Union nor the granting authority can be held responsible for them. This work has received funding from the European Research Council (ERC) under the European Union’s Horizon 2020 research and innovation program (Grant agreement No.\ 945806) and is supported by the Deutsche Forschungsgemeinschaft (DFG, German Research Foundation) under Germany’s Excellence Strategy EXC 2181/1-390900948 (the Heidelberg STRUCTURES Excellence Cluster).
\end{acknowledgements}

\bibliography{astrofritz_edited} 

\clearpage
\begin{appendix}
\twocolumn[
\begin{@twocolumnfalse}
\section*{\centering Appendices}
\end{@twocolumnfalse}
]
\section{Kinetic energy density evolution} \label{sec:appendixA}
\begin{figure}
    \includegraphics[width=\linewidth]{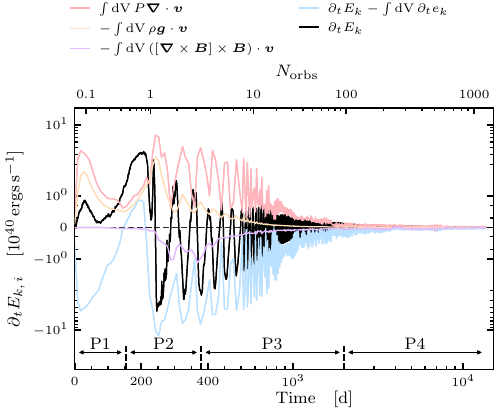}
    \caption{Same as Fig.~\ref{fig:emag_evolution} but for the kinetic energy density evolution equation (Eq.~\ref{eq:ekin_evolution}). Since we are integrating over the entire domain with periodic boundary conditions, we leave out the advection term and plotted is the derivative of the total kinetic energy of the gas $\partial_t E_\mathrm{k}$ (black), the integrated pressure-volume work term $\int \mathrm{dV}\, P \boldsymbol{\nabla}\cdot \boldsymbol{v}$ (red), the gravitational term $\int \mathrm{dV}\, \rho \boldsymbol{g}\cdot \boldsymbol{v}$} (yellow), the Lorentz term $\int \mathrm{dV}\, ([ \boldsymbol{\nabla\times B}]\times\boldsymbol{B})\cdot \boldsymbol{v}$ (violet) and the dissipation term as the residual of the equation $\partial_t E_\mathrm{k} - \int \mathrm{dV} \partial_t \rho e_\mathrm{k}$ (blue).
    \label{fig:ekin_evolution}
\end{figure}

Similarly to the magnetic energy density evolution equation in Eq.~(\ref{eq:emag_eq}), one can derive the kinetic energy density evolution equation by multiplying the MHD Euler equation with the velocity and utilizing the continuity equation. The resulting equation reads,
\begin{equation}
\begin{aligned}[b]
    \partial_t (\rho e_\mathrm{k}) + \boldsymbol{\nabla} \cdot (\rho e_\mathrm{k}\boldsymbol{v}) = &- \boldsymbol{\nabla} \cdot (P\boldsymbol{v}) + P \boldsymbol{\nabla} \cdot \boldsymbol{v} + \rho\boldsymbol{g}\cdot \boldsymbol{v} \\ &+ \frac{1}{4\pi}[(\boldsymbol{\nabla} \times \boldsymbol{B}) \times \boldsymbol{B}]\cdot \boldsymbol{v},
\end{aligned}\label{eq:ekin_evolution}
\end{equation}
with $e_\mathrm{k} = \vert\boldsymbol{v}\vert^2 /2$, $P$, $\boldsymbol{g}$ are the specific kinetic energy, pressure, and gravitational acceleration, respectively. Again, we integrate the evolution equation over the entire domain, which yields the contributions from the different terms in Fig.~\ref{fig:ekin_evolution}. From left to right, the two terms on the left-hand side in Eq.~(\ref{eq:ekin_evolution}) represent the time derivative of the kinetic energy density and its advection, while the right hand-side accounts for work done by pressure force, (de-) compression (pressure-volume work), gravity and the Lorentz force. The advection and pressure force term can again be ignored, given the periodicity of our domain. As expected, during the first orbit of the companion, orbital energy is directly transferred into the kinetic energy budget of the gas (Fig.~\ref{fig:ekin_evolution}), but the main contribution appears to come from pressure-volume work ($P\boldsymbol{\nabla} \cdot \boldsymbol{v}$), most likely caused by the perturbed hydrostatic equilibrium of the primary star and decompression in the spiral arms. After the first orbit $t\, {>} \, 200 \, \mathrm{d}$, all terms (except the Lorentz contribution) start to oscillate with the orbital frequency of the binary and dissipation (blue line) is compensating the energy gain by pressure and gravitational work. The magnetic fields start to back react on the fluid flow and contributing as a sink of kinetic energy for the rest of the simulation. As the simulation evolves and the entire envelope is becoming ejected, the change in kinetic energy starts to flatten and converge to zero.

\section{Measuring the outflow rates in the simulation involving the neutron star companion} \label{sec:appendixB}
\begin{figure}
    \includegraphics[width=\linewidth]{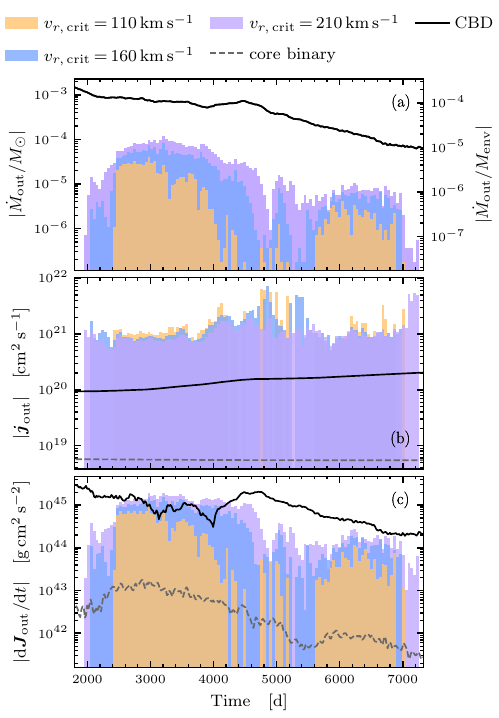}
    \caption{Same as Fig.~\ref{fig:outflow_binned_BH} but for the simulation involving the NS companion (see Sect.~\ref{sec:ns_comp}). For more information, see Sect.~\ref{sec:jet_quants}.}
    \label{fig:outflow_binned_NS}
\end{figure}

Identical to the procedure in Sect.~\ref{sec:jet_quants}, we utilize tracer particles to measure the ejection rates through the magnetized outflows in the simulation involving a NS (see Sect.~\ref{sec:ns_comp}). The findings are summarized in Table~\ref{tab:quant_jet_out_NS} and Fig.~\ref{fig:outflow_binned_NS}.  

\begin{table}
\centering
\caption{Same as Table~\ref{tab:quant_jet_out} but for the neutron star companion (see Text in Sect.~\ref{sec:ns_comp}).}
	\label{tab:quant_jet_out_NS}
    \begin{tabular}{l|ccc}
	\hline\hline
    Quantity & \multicolumn{3}{c}{Values}\\ 
	$v_\mathrm{r, crit} \quad \mathrm{km \, s^{-1}}$ &110&160&210 \\
	\hline
	$\langle v_\mathrm{r}\rangle_\mathrm{p} \quad [\mathrm{km \, s^{-1}}]$                   &164	&193 & 224 \\
    $\langle v_\mathrm{r}\rangle_\mathrm{t} \quad [\mathrm{km \, s^{-1}}]$                   &155	&186 & 218 \\
    $\langle v_r/v_\mathrm{orb}\rangle_\mathrm{p}$                                         &0.74 &0.85 &0.99\\
    $\langle v_r/v_\mathrm{orb}\rangle_\mathrm{t}$                                         &0.69 &0.83 &0.97\\ 
	$N_\mathrm{out, \,tot}$	                                                               &21,758	& 13,209 & 5,337 \\
	$M_\mathrm{out, \, tot}\quad [M_\odot]$                                                  &0.18&0.11&0.04  \\
	$M_\mathrm{out, tot} \quad [M_\mathrm{env}]$                                             &0.027&0.017&0.007\\
    $\langle M_\mathrm{out}\rangle_\mathrm{t} \quad [10^{-3}\, {\times}\, M_\odot]$           &1.04&0.60&0.20 \\ 
	$\langle \dot{M}_\mathrm{out}\rangle_t \quad [\mathrm{10^{-5}\, M_\odot \, d^{-1}}]$	   &2.3 &1.3& 0.44\\
    $\langle |\boldsymbol{j}_\mathrm{out}|\rangle_\mathrm{t} \quad \mathrm{10^{21}\, cm^{2}\, s^{-1}}$                              &1.16 &1.42&1.58\\   
    $\langle |\boldsymbol{j}_\mathrm{out}| /j(r\,{=}1.1a)\rangle_\mathrm{t}$               &43.1 &53.2&59.14\\                
	$\langle \mathrm{d} \boldsymbol{J}_\mathrm{out} /\mathrm{d}t\rangle_\mathrm{t}\quad [\mathrm{10^{44}\,g\ cm^2\, s^{-2}}]$       &4.8& 3.1&1.1\\
	\hline
	\end{tabular}	
\end{table}

\section{Movies of the CE interactions}\label{sec:appendix_movies}

\begin{table*}
    \centering
    \setlength{\tabcolsep}{8.2pt}
    \caption{Time evolution movies of different quantities for the Hydro and MHD CE simulation.}
    \begin{tabular}{c|cc}
         \hline\hline
         ID & File name & \makecell{Description\\}\\
         \hline
         M1 & \href{https://doi.org/10.5281/zenodo.14967081}{movie\_density.mov}& \makecell{Density evolution for the MHD (left column) and\\ Hydro simulation (right column). The first and second row \\ shows the face-on  ($x \, \text{--}\, y$) and edge-on ($x \, \text{--}\, z$) view, respecively.}\\
         \hline
         M2 & \href{https://doi.org/10.5281/zenodo.14986851}{movie\_vr.mov}& \makecell{Radial velocity in the edge-on ($x$--$z$) view. The MHD secnario\\ is shown in (a), while in (b) the Hydro simulation is shown.}\\
         \hline
         M3 & \href{https://doi.org/10.5281/zenodo.14966754}{movie\_bfld.mov}
         & \makecell{Time evolution of the absolute magnetic-field strength \\in the edge-on ($x \, \text{--}\, z$) and face-on ($x \, \text{--}\, y$) view. The \\individual slices are line-integrated convolution plots.}\\
         \hline
         \hline
    \end{tabular}
    \tablefoot{In M1 and M2, we show the density and radial velocity evolution for both runs, respectively. The evolution of the magnetic field strength is shown in M2. The primary core and the companion are marked with a \qq{x} and \qq{+,} respectively.}
    \label{tab:movietable}
\end{table*}

In Table~\ref{tab:movietable} we summarize the movies presented in this work. There we show the temporal evolution of the density (M1), radial velocity (M2) and absolute magnetic-field strength (M3).

\end{appendix}
 
\end{document}